\documentclass[fleqn,usenatbib]{mnras}

\usepackage{newtxtext,newtxmath}

\usepackage[T1]{fontenc}
\usepackage{adjustbox}

\usepackage{graphicx}	
\usepackage{subfig}
\usepackage{pdflscape}
\usepackage{afterpage}

\DeclareCaptionLabelFormat{continued}{Figure #1~#2 (Continued)}
\title[Tidal tails in Open Clusters]{A \textit{Gaia} EDR3 search for tidal tails in disintegrating open clusters}

\author[S. Bhattacharya et al.]{
Souradeep Bhattacharya,$^{1}$\thanks{E-mail: souradeep@iucaa.in}
Khushboo K. Rao,$^{2}$
Manan Agarwal,$^{3}$
Shanmugha Balan,$^{2}$
and Kaushar Vaidya$^{2}$
\\
$^{1}$Inter University Centre for Astronomy and Astrophysics, Ganeshkhind, Post Bag 4, Pune 411007, India\\
$^{2}$Department of Physics, Birla Institute of Technology and Science – Pilani, Rajasthan 333031, India\\
$^{3}$Department of Physics and Kavli Institute for Astrophysics and Space Research, Massachusetts Institute of Technology, Cambridge, MA 02139, USA
}

\date{Accepted October 5, 2022. Received October 5, 2022; in original form December 23, 2021}

\pubyear{2022}

\begin{document}
\label{firstpage}
\pagerange{\pageref{firstpage}--\pageref{lastpage}}
\maketitle

\begin{abstract}
We carry out a search for tidal tails in a sample of open clusters with known relatively elongated morphology. We identify the member stars of these clusters from the precise astrometric and deep photometric data from \textit{Gaia} Early Data Release 3 using the robust membership determination algorithm, ML-MOC. We identify 46 open clusters having a stellar corona beyond the tidal radius, 20 of which exhibit extended tails aligned with the cluster orbit direction in galactocentric coordinates. Notably we find NGC~6940 (at a distance of $\sim1$~kpc) is the furthest open cluster exhibiting tidal tails that are $\sim50$~pc from its center, while also identifying $\sim40$~pc long tidal tails for the nearby Pleiades. Using the minimum spanning tree length for the most massive stars relative to all cluster members, we obtain the mass segregation ratio ($\rm\lambda_{MSR}$) profiles as a function of the number of massive stars in each cluster. From these profiles, we can classify the open clusters into four classes based on the degree of mass segregation experienced by the clusters. {We find that clusters in the most mass segregated classes are the oldest on average and have the flattest mass function slope.} Of the 46 open clusters studied in this work, 41 exhibit some degree of mass segregation. Furthermore, we estimate the initial masses (M$\rm_{i}$) of these open clusters finding that some of them, having M$\rm_{i}\gtrsim 10^{4} M_{\sun}$, could be the dissolving remnants of Young Massive Clusters.
\end{abstract}




\section{Introduction}
\label{sect:intro}

All stars likely form in clusters, which over time disintegrate to make up the field population of the Galaxy \citep{Zwart10}. Two-body relaxation over time leads to more-massive stars sinking to the central regions of the cluster while less-massive stars occupy a larger volume and gradually evaporate. 
Additionally, the Galactic potential perturbs clusters to form tidal tails, an effect observed for both open (e.g. Berkeley~17 - \citealt{Chen04,Bhattacharya17b}; NGC~6791 - \citealt{Dalessandro15}) and globular clusters (e.g. Palomar~5 - \citealt{Odenkirchen01}; NGC~5466 - \citealt{Belokurov06}). 

Identification of such tidal features in open clusters, which are more abundant in the Galactic disc, has long been challenging from photometric observations alone, despite the use of novel techniques \citep[e.g.][]{Bhattacharya17a}. This is because low-density features such as tidal tails are difficult to identify against the strong field contamination in the Galactic disc. Reliable astrometry (proper motion and parallax) from \textit{Gaia} Data Release 2 \citep[\textit{Gaia} DR2;][]{Gaia18} and recently \textit{Gaia} Early Data Release 3 \citep[\textit{Gaia} EDR3;][]{Gaia21} has allowed for improved field-star decontamination of many nearby open clusters, resulting in the recent discovery of a many open clusters with extended stellar coronae \citep[e.g.][]{Carrera19,Meingast21,Tarricq22}, as well as 13 of those with tidal tails -- Hyades: \citet{Roser19,Meingast19,Jerabkova21}; Coma~Berenices: \citet{Tang19}; Ruprecht 147: \citet{Yeh19}; Praesepe: \citet{Roser19b}; NGC~2506: \citet{Gao20}; Blanco~1: \citet{Zhang20}; M~67: \citet{Gao20b}; UBC~274: \citet{Castro20}; Alpha Persei: \citet{Nikiforova20}; NGC~2516, NGC~6633: \citet{Pang21}; NGC~752: \citet{Bhattacharya21}; Pleiades: \citet{Li21}. \citet{Tarricq22} recently carried out a systematic search for tidal tails in open clusters by identifying elongated features in the 2D projected spatial distribution of cluster members identified from \textit{Gaia} EDR3 data. While they identified elongated features in 72 open clusters, terming them as tidal tails, {they did not check the alignment of these extended features with the cluster orbit direction.} 

In this work, we utilise the 2D shape morphology computed by \citet{Hu21} to systematically search for tidal features in a sample of open clusters showing elongated morphology. The cluster selection and membership determination using \textit{Gaia} EDR3 is presented in Section~\ref{sect:data}. {As in \citet{Bhattacharya21} for NGC~752, we adopt the following criteria for the 2D projected extended spatial features of any cluster to be tidal tails: in tri-dimensional projections in the galactocentric coordinate system, the cluster members constituting the leading (trailing) tails, identified in the 2D projected spatial distribution, should point towards (away from) a direction intermittent between the direction of the cluster orbit and the Galactic Center (GC) in at least two of the three planes.} The methodology for reliably determining tidal features is presented in Section~\ref{sect:analysis}. We discuss the results, including a discussion on the mass segregation of the clusters, in Section~\ref{sect:disc} and conclude in Section~\ref{sect:con}.

\section{Cluster and membership selection}
\label{sect:data}

\subsection{Cluster selection}
\label{sect:clus_sel}

\begin{figure}
	\includegraphics[width=\columnwidth]{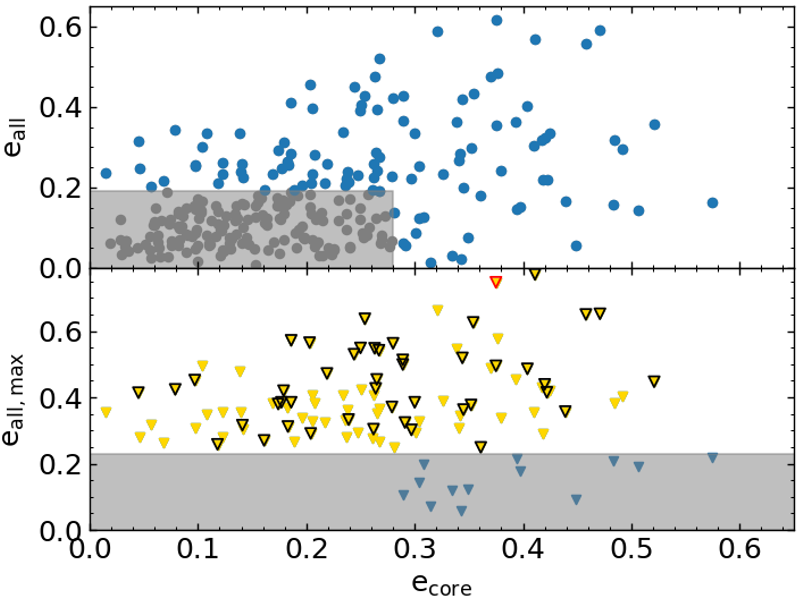}
    \caption{[Top] e$\rm_{core}$ vs e$\rm_{all}$ for 265 open clusters from \citet{Hu21}. Those 106 open clusters marked in blue have been classified as being the most elongated while those in the grey shaded region have been discarded. [Bottom] e$\rm_{core}$ vs e$\rm_{all, max}$ for the 106 open clusters selected above. Those in the grey shaded region are discarded while those shown in yellow have been searched for extended features. {NGC~752 studied in \citet{Bhattacharya21} is marked in red.} The 45 open clusters showing extended features in this work have been enclosed in black.}
    \label{fig:clus_sel}
\end{figure}

Using cluster membership determined by \citet{cantat18} from \textit{Gaia} DR2 proper motion and parallax measurements of individual stars, \citet{Hu21} obtained shape parameters for 265 open clusters, particularly including e$\rm_{core}$ (core ellipticity) and e$\rm_{all}$ (all member ellipticity). We posit that the open clusters showing the most elongated morphology (evidenced from their e$\rm_{core}$ and e$\rm_{all}$) are more likely to harbour tidal tails which may be identified with improved membership determination using the more accurate \textit{Gaia} EDR3 data. We note that only 7 of the 13 known open clusters with tidal tails are present in the catalogue of \citet{Hu21}, namely Praesepe, M~67, Alpha~Persei, NGC~2516, NGC~6633, NGC~752 and Pleiades. In order to shortlist clusters with possible tidal tails, we carried out the following steps:
\begin{enumerate}
    \item We first selected a third of the open clusters having the highest e$\rm_{all}$ values, since these are the most elongated ones. This corresponded to open clusters having e$\rm_{all} > 0.19$. Given the cluster outskirts may not have been adequately covered by the members identified using \textit{Gaia} DR2, any cluster showing elongated core morphology (top 20\%) were further selected, corresponding to e$\rm_{core} > 0.28$. Figure~\ref{fig:clus_sel} [top] shows the distribution of e$\rm_{all}$ values against the e$\rm_{core}$ values determined by \citet{Hu21} for 265 open clusters. The 106 clusters satisfying the aforementioned criteria are marked in blue, while the others are in grey.
    \item {For each of these 106 clusters, we determined the maximum ellipticity, e$\rm_{all,max}$, as the upper limit of e$\rm_{all}$.} The bottom 10\% of clusters which had least e$\rm_{all,max}$ values, making them unlikely to be elongated in the outskirts, were discarded. We were thus left with 95 cluster candidates which may be searched for possible tidal tails. Figure~\ref{fig:clus_sel} [bottom] shows the distribution of e$\rm_{all, max}$ values against the e$\rm_{core}$ for these 95 clusters shown in yellow, while those in blue were discarded. 
    \item Of the 95 clusters, only three with already known tidal tails survive the selection (Alpha Persei, NGC~752 and the Pleiades). The Pleiades has only short tidal tails discovered \citep{Ye21} with possibility of longer tails being uncovered. {Alpha Persei's tidal tails were only identified by \citet{Nikiforova20} as photometric overdensities against the forground and background sources, and the tail members were not identified from kinematic measurements. We thus continue to include Alpha Persei and the Pleiades in our study.} {NGC~752 already has long tails discovered and studied in depth in \citet{Bhattacharya21}.} {It} is marked in red in Figure~\ref{fig:clus_sel} [bottom] and {has} been excluded from this study. The other {94} clusters were chosen for membership determination and a search for tidal tails was carried out in some of them (see Section~\ref{sect:analysis}). 
\end{enumerate}

\subsection{Membership determination}
\label{sect:clus_data}

\begin{figure*}
	\includegraphics[width=\textwidth]{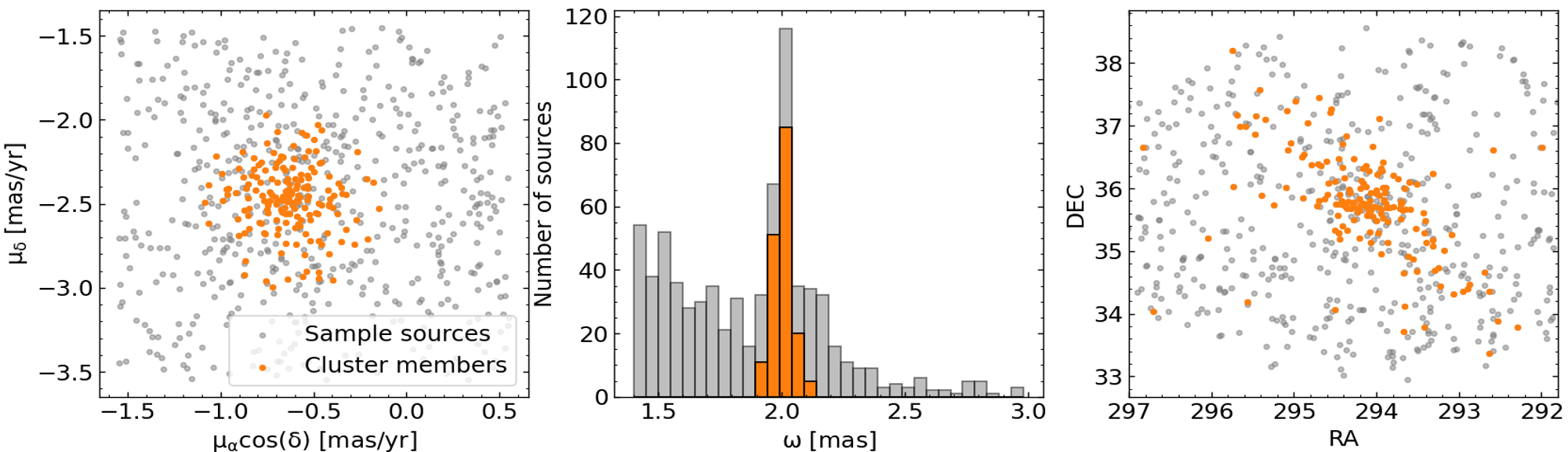}
    \caption{The proper-motion, parallax and spatial distributions of the \textit{Sample sources} and the cluster members of Teutsch~35 identified by ML-MOC.}
    \label{fig:mlmoc}
\end{figure*}

\begin{figure*}
	\includegraphics[width=\textwidth]{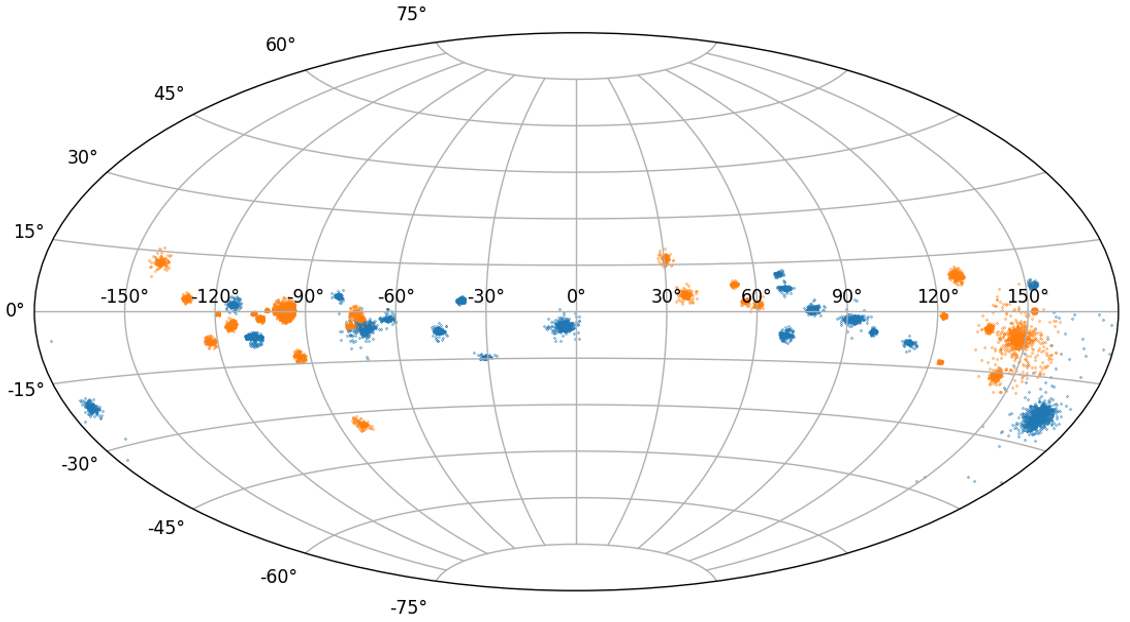}
    \caption{2D projection of identified cluster members of the {46} clusters with extended features in Galactic coordinates. Those shown in orange only show an extended corona while those in blue also show tidal tails.}
    \label{fig:gal}
\end{figure*}

For each of the 94 clusters, we obtain \textit{Gaia} EDR3 positions, trigonometric parallaxes ($\rm\omega$), proper motions (PM), radial velocity (RV) as well as photometry in three broad-band filters (G, BP, and RP) within a search radius of twice the entire cluster half-length around its center as noted by \citet{Hu21}. {Then quality cuts are applied as mentioned in \citet[][see their Section 3.1]{Agarwal21}.} This sample is termed \textit{All sources} for each cluster. We use ML-MOC \citep{Agarwal21} to identify members for each cluster using the proper motion and parallax information. ML-MOC is based on the k-nearest neighbour algorithm \citep[kNN;][]{kNN} and the Gaussian mixture model \citep[GMM;][]{GMM}. It identifies cluster members in the PM--$\rm\omega$ parameter space, independent of the spatial density of the cluster. This allows for the identification of faint extended spatial structures, as already demonstrated for NGC~752 \citep{Bhattacharya21}. Note the completeness of \textit{Gaia} EDR3 data and the contamination fraction for ML-MOC identified open cluster members are discussed in Appendix~\ref{sect:comp}. Briefly, the \textit{Gaia} EDR3 data for open clusters is more than 90\% complete down to G~$\sim$19.5~mag with a contamination $\sim2.3\%$. Details of the membership selection are described in \citet{Agarwal21} and we briefly describe it as follows, taking the cluster Teutsch~35 as an example:
\begin{enumerate}
    \item In the first step of the membership selection, kNN is utilised to remove the obvious field stars such that the remaining member candidates have more cluster members than field stars. This is done by applying kNN (with 25 nearest neighbours) to those stars within the central region of the cluster (within a radius equal to the cluster core half-length noted by \citealt{Hu21}) to calculate a broad range of PM and $\omega$ values for the cluster members. For Teutsch~35, we find the range of PM in RA, $\rm\mu_{\alpha}cos(\delta)=$ $-1.546$ -- $0.549$ mas/yr, PM in DEC, $\mu_{\delta}=$ $-3.543$ -- $-1.453$ mas/yr, and $\omega=$ $1.401$ -- $2.983$. The stars in \textit{All sources} which fall in the aforementioned parameter ranges are termed \textit{Sample sources}. Their spatial, PM and $\omega$ distributions are shown in Figure~\ref{fig:mlmoc}.
    \item For the second step, a three dimensional GMM is used in the PM--$\omega$ parameter space of the \textit{Sample sources} to distinguish between the cluster and field members, also assigning a membership probability ($\rm p_{memb}$). Those sources having $\rm p_{memb}\geq0.6$ are considered as high probability members. 
    \item A small number of stars having $\rm 0.2 \leq p_{memb}\leq0.6$ are also considered as members if their $\rm\omega$ values lie within the range of $\rm\omega$ values specified by the members that have $\rm p_{memb}\geq0.8$. This results in the identification of 172 members (N$\rm_{memb}$) in Teutsch~35 whose spatial, PM and $\omega$ distributions are also shown in Figure~\ref{fig:mlmoc}. Note the extended tails are clearly {visible} in the spatial distribution. 
\end{enumerate}

Among the 94 clusters whose membership determination was attempted, ML-MOC failed to securely identify cluster members in 20 of them for the large search radius required to find extended structures. This was because the \textit{Sample sources} of these clusters had overlapping cluster members and field stars in the PM--$\omega$ parameter space, not allowing for two distinguishable gaussian fits \citep[see][for details of such cases]{Agarwal21}. While ML-MOC may have been more successful in identifying members in a smaller search radius, such a search was not attempted as we are interested primarily in extended features in the cluster outskirts. A further 17 clusters had very few members determined (also in the \textit{Gaia} DR2 membership determination by \citealt{cantat18}) and were discarded as it would not have been possible to securely identify extended features in clusters with such low member density. Finally, 11 more clusters did not have members identified beyond their tidal radius (fitted in Section~\ref{sect:tr}) and were also excluded from further study. The remaining 46 clusters with extended features have been studied here and searched for tidal tails. These also include 33 open clusters studied by \citet{Tarricq22}, 10 of which were identified as having extended features from their 2D spatial distribution but the tidal nature of their tails is yet to be determined.

\afterpage{%
    \thispagestyle{empty}
    \begin{landscape}
    \captionof{table}{Cluster parameters for the 46 open clusters with extended features. Column 1: Name; Columns 2--3: Equatorial coordinates; Columns 4--7: Parameters for isochrone fit from \citet{Bossini19}; Columns 8--9: Cluster ellipticity from \citet{Hu21}; Column 10: Number of cluster members identified; Column 11--13: Mean parallax and proper motions of identified members; Column 14: Number of identified members with \textit{Gaia} EDR3 RV measurements; Column 15: Mean RV of identified members; Column 16: The extent of the cluster corona; Column 17: Photometric cluster mass; Column 18: Tidal radius; Column 19: Initial cluster mass.}
    \label{table:clus}
        \centering 
        \adjustbox{max width=1.35\textwidth}{
        \begin{tabular}{cccccccccccccccccccc}
            \hline
            Name & RA & DEC & Age & [Fe/H] & A$\rm_{V}$ & Dist & e$\rm_{core}$ & e$\rm_{all}$ & N$\rm_{memb}$ & $\rm \omega$ & $\rm \mu_{\alpha}cos(\delta)$ & $\rm \mu_{\delta}$ & N$\rm_{RV}$ & RV & R$\rm_{C}$  & m$\rm_{p}$ & r$\rm_{t}$ & M$\rm_{i}$ \\
            & deg & deg & log(yr) & & mag & pc & & & & mas & mas/yr & mas/yr & & km/s & pc & M$_{\sun}$ & pc & M$_{\sun}$ \\
            \hline
Alessi 1 & 13.343 & 49.536 & 8.935 & 0.0 & 0.322 & 709.58 & 0.422 $\pm$ 0.033 & 0.219 $\pm$ 0.196 & 59 & 1.409 $\pm$ 0.007 & 6.486 $\pm$ 0.012 & -6.429 $\pm$ 0.021 & 10 & -2.14 $\pm$ 5.03 & 12.05 & 75.47 & 5.49 & 5518$^{+3964}_{-222}$  \\
Alessi 2 & 71.602 & 55.199 & 8.759 & 0.0 & 0.5 & 580.23 & 0.28 $\pm$ 0.026 & 0.422 $\pm$ 0.141 & 201 & 1.625 $\pm$ 0.004 & -0.925 $\pm$ 0.011 & -1.1 $\pm$ 0.01 & 22 & -9.87 $\pm$ 2.43 & 16.37 & 183.47 & 7.35 & 3440$^{+2156}_{-1236}$ \\
Alessi 24 & 260.764 & -62.693 & 7.969 & -0.154 & 0.363 & 457.72 & 0.411 $\pm$ 0.008 & 0.568 $\pm$ 0.203 & 149 & 2.066 $\pm$ 0.004 & -0.436 $\pm$ 0.011 & -8.975 $\pm$ 0.013 & 10 & 11.67 $\pm$ 1.82 & 31.31 & 118.9 & 6.63 & 397$^{+140}_{-87}$ \\
Alessi 5 & 160.819 & -61.081 & 7.723 & -0.133 & 0.592 & 350.11 & 0.174 $\pm$ 0.007 & 0.293 $\pm$ 0.087 & 221 & 2.52 $\pm$ 0.003 & -15.401 $\pm$ 0.015 & 2.55 $\pm$ 0.018 & 21 & 10.03 $\pm$ 1.79 & 20 & 205.25 & 8.23 & 390$^{+76}_{-49}$ \\
Alessi 6 & 220.058 & -66.127 & 8.803 & -0.382 & 0.74 & 813.58 & 0.097 $\pm$ 0.01 & 0.256 $\pm$ 0.196 & 141 & 1.128 $\pm$ 0.002 & -10.56 $\pm$ 0.01 & -5.515 $\pm$ 0.011 & 4 & -12.16 $\pm$ 1.57 & 19.53 & 170.56 & 7.5 & 3902$^{+2517}_{-1436}$\\
Alessi 62 & 284.026 & 21.597 & 8.997 & 0.0 & 0.603 & 575.17 & 0.265 $\pm$ 0.03 & 0.394 $\pm$ 0.06 & 201 & 1.613 $\pm$ 0.004 & 0.269 $\pm$ 0.009 & -1.146 $\pm$ 0.011 & 12 & 13.35 $\pm$ 1.45 & 14.55 & 179.37 & 7.6 & 7453$^{+5127}_{-2894}$ \\
Alessi Teutsch 3 & 118.228 & -53.022 & 8.2 & 0.0 & 0.439 & 668.65 & 0.186 $\pm$ 0.027 & 0.285 $\pm$ 0.1 & 192 & 1.407 $\pm$ 0.004 & -5.334 $\pm$ 0.013 & 8.857 $\pm$ 0.021 & 8 & 15.65 $\pm$ 1.22 & 9.88 & 167.81 & 6.16 & 748$^{+312}_{-189}$ \\
Alpha Persei & 51.597 & 48.921 & 7.938 & 0.14 & 0.055 & 146.15 & 0.234 $\pm$ 0.014 & 0.338 $\pm$ 0.068 & 476 & 5.73 $\pm$ 0.008 & 22.782 $\pm$ 0.054 & -25.442 $\pm$ 0.049 & 152 & -0.77 $\pm$ 1.27 & 36.27 & 554.12 & 10.62 & 1026$^{+118}_{-182}$ \\
ASCC 101 & 288.399 & 36.369 & 8.694 & 0.0 & 0.053 & 376.36 & 0.354 $\pm$ 0.027 & 0.432 $\pm$ 0.194 & 103 & 2.515 $\pm$ 0.006 & 1.022 $\pm$ 0.024 & 1.215 $\pm$ 0.025 & 19 & -18.56 $\pm$ 4.54 & 11.089 & 82.84 & 5.69 & 2448$^{+1632}_{-926}$ & \\
ASCC 41 & 116.674 & 0.137 & 8.04 & 0.0 & 0.053 & 278.74 & 0.404 $\pm$ 0.02 & 0.401 $\pm$ 0.085 & 192 & 3.371 $\pm$ 0.008 & 0.637 $\pm$ 0.021 & -3.946 $\pm$ 0.019 & 22 & -10.06 $\pm$ 2.54 & 19.67 & 102.63 & 6.09 & 425$^{+172}_{-104}$ \\
BH 164 & 222.311 & -66.465 & 7.7 & 0.0 & 0.381 & 396.46 & 0.186 $\pm$ 0.039 & 0.412 $\pm$ 0.16 & 286 & 2.373 $\pm$ 0.004 & -7.43 $\pm$ 0.019 & -10.691 $\pm$ 0.015 & 16 & 1.84 $\pm$ 1.34 & 20.45 & 206.42 & 7.83 & 382$^{+71}_{-46}$\\
BH 99 & 159.553 & -59.168 & 7.908 & 0.0 & 0.203 & 413.43 & 0.254 $\pm$ 0.017 & 0.427 $\pm$ 0.21 & 443 & 2.235 $\pm$ 0.002 & -14.48 $\pm$ 0.012 & 0.977 $\pm$ 0.015 & 31 & 12.28 $\pm$ 1.07 & 17.66 & 344.68 & 9.55 & 701$^{+147}_{-95}$\\
Collinder 463 & 27.031 & 71.738 & 8.5 & 0.0 & 0.836 & 805.38 & 0.219 $\pm$ 0.028 & 0.259 $\pm$ 0.213 & 549 & 1.151 $\pm$ 0.001 & -1.739 $\pm$ 0.005 & -0.375 $\pm$ 0.007 & 11 & -10.53 $\pm$ 0.66 & 35.55 & 678.02 & 11.2 & 2688$^{+1026}_{-628}$ \\
Gulliver 21 & 106.961 & -25.462 & 8.472 & 0.0 & 0.187 & 622.87 & 0.289 $\pm$ 0.016 & 0.367 $\pm$ 0.132 & 129 & 0.741 $\pm$ 0.002 & -0.233 $\pm$ 0.006 & 0.652 $\pm$ 0.006 & 14 & 41.2 $\pm$ 0.99 & 21.57 & 206.06 & 7.42 & 1552$^{+794}_{-469}$ \\
Gulliver 36 & 123.185 & -35.111 & 8.8 & 0.0 & 0.83 & 1185.77 & 0.291 $\pm$ 0.02 & 0.055 $\pm$ 0.269 & 220 & 1.538 $\pm$ 0.004 & -1.913 $\pm$ 0.011 & 4.229 $\pm$ 0.011 & 3 & 10.22 $\pm$ 0.69 & 18.6 & 138.44 & 6.75 & 3737$^{+2463}_{-140}$ &\\
Gulliver 44 & 127.249 & -38.095 & 9.182 & 0.0 & 1.1 & 1258.92 & 0.279 $\pm$ 0.027 & 0.229 $\pm$ 0.142 & 146 & 0.803 $\pm$ 0.003 & -0.653 $\pm$ 0.008 & 2.365 $\pm$ 0.01 & 4 & 14.44 $\pm$ 1.16 & 12.73 & 173.58 & 7.46 & 14108$^{+10181}_{-5698}$ \\
Haffner 13 & 115.209 & -30.073 & 7.497 & 0.0 & 0.168 & 533.09 & 0.521 $\pm$ 0.057 & 0.356 $\pm$ 0.091 & 471 & 1.787 $\pm$ 0.003 & -6.173 $\pm$ 0.013 & 5.921 $\pm$ 0.009 & 8 & 36.97 $\pm$ 0.38 & 18.42 & 325.8 & 9.03 & 466$^{+47}_{-31}$ \\
IC 2602 & 160.613 & -64.426 & 7.547 & 0.0 & 0.096 & 152.34 & 0.079 $\pm$ 0.052 & 0.343 $\pm$ 0.081 & 437 & 6.61 $\pm$ 0.007 & -17.757 $\pm$ 0.061 & 10.654 $\pm$ 0.057 & 51 & 16.4 $\pm$ 1.29 & 25.86 & 259.76 & 8.77 & 399$^{+50}_{-33}$\\
IC 4665 & 266.554 & 5.615 & 7.636 & -0.03 & 0.43 & 315.5 & 0.25 $\pm$ 0.049 & 0.39 $\pm$ 0.159 & 246 & 2.866 $\pm$ 0.01 & -0.836 $\pm$ 0.018 & -8.505 $\pm$ 0.021 & 18 & -12.4 $\pm$ 1.88 & 16.64 & 159.79 & 7.4 & 295$^{+55}_{-36}$\\
IC 4756 & 279.649 & 5.435 & 8.987 & 0.0 & 0.397 & 478.85 & 0.204 $\pm$ 0.04 & 0.211 $\pm$ 0.08 & 451 & 2.105 $\pm$ 0.002 & 1.277 $\pm$ 0.01 & -4.982 $\pm$ 0.011 & 50 & -25.53 $\pm$ 1.86 & 32.94 & 461.76 & 10.69 & 8353$^{+5183}_{-2977}$\\
M39 & 322.889 & 48.247 & 8.491 & 0.0 & 0.0 & 281.97 & 0.458 $\pm$ 0.023 & 0.556 $\pm$ 0.094 & 308 & 3.368 $\pm$ 0.005 & -7.391 $\pm$ 0.022 & -19.711 $\pm$ 0.028 & 26 & -5.94 $\pm$ 1.68 & 28.26 & 188.2 & 7.38 & 1585$^{+840}_{-494}$ & -\\
NGC 225 & 10.805 & 61.774 & 8.254 & 0.0 & 0.821 & 707.62 & 0.3 $\pm$ 0.083 & 0.334 $\pm$ 0.051 & 136 & 1.444 $\pm$ 0.005 & -5.339 $\pm$ 0.01 & -0.176 $\pm$ 0.012 & 3 & 11.45 $\pm$ 0.93 & 12.05 & 135.4 & 6.57 & 775$^{+362}_{-216}$\\
NGC 1039 & 40.531 & 42.722 & 8.101 & 0.0 & 0.239 & 513.33 & 0.141 $\pm$ 0.009 & 0.258 $\pm$ 0.058 & 705 & 2.004 $\pm$ 0.003 & 0.658 $\pm$ 0.008 & -5.783 $\pm$ 0.01 & 50 & -6.38 $\pm$ 1.03 & 18.01 & 589.24 & 11.77 & 1292$^{+295}_{-188}$\\
NGC 1528 & 63.878 & 51.218 & 8.596 & 0.0 & 0.802 & 990.38 & 0.177 $\pm$ 0.019 & 0.247 $\pm$ 0.138 & 385 & 0.969 $\pm$ 0.002 & 2.146 $\pm$ 0.007 & -2.298 $\pm$ 0.007 & 3 & -7.43 $\pm$ 0.48 & 14.26 & 464.13 & 10.75 & 2822$^{+1330}_{-794}$\\
NGC 1662 & 72.198 & 10.882 & 8.957 & -0.11 & 0.615 & 369.83 & 0.244 $\pm$ 0.016 & 0.451 $\pm$ 0.079 & 275 & 2.434 $\pm$ 0.005 & -1.135 $\pm$ 0.013 & -1.972 $\pm$ 0.012 & 34 & -8.25 $\pm$ 2.18 & 20.47 & 217.21 & 7.82 & 6683$^{+4462}_{-2532}$\\
NGC 1901 & 79.561 & -68.294 & 8.918 & -0.08 & 0.146 & 409.07 & 0.267 $\pm$ 0.166 & 0.521 $\pm$ 0.021 & 156 & 2.373 $\pm$ 0.005 & 1.67 $\pm$ 0.042 & 12.655 $\pm$ 0.019 & 22 & 3.49 $\pm$ 2.71 & 19.64 & 131.48 & 6.45 & 5497$^{+3787}_{-2137}$\\
NGC 2423 & 114.299 & -13.863 & 9.07 & 0.08 & 0.19 & 930.25 & 0.179 $\pm$ 0.015 & 0.313 $\pm$ 0.107 & 445 & 1.06 $\pm$ 0.002 & -0.751 $\pm$ 0.006 & -3.579 $\pm$ 0.006 & 33 & 22.75 $\pm$ 0.97 & 26.8 & 499.15 & 10.78 & 10993$^{+7023}_{-4014}$\\
NGC 2448 & 116.034 & -24.837 & 7.997 & 0.0 & 0.119 & 1077.95 & 0.361 $\pm$ 0.042 & 0.179 $\pm$ 0.07 & 144 & 0.899 $\pm$ 0.003 & -3.392 $\pm$ 0.006 & 2.951 $\pm$ 0.007 & 3 & 40.43 $\pm$ 2.65 & 15.07 & 193.0 & 7.469 & 546$^{+168}_{-105}$ \\
NGC 2451B & 116.128 & -37.954 & 7.604 & 0.0 & 0.232 & 335.89 & 0.345 $\pm$ 0.057 & 0.199 $\pm$ 0.164 & 424 & 2.756 $\pm$ 0.004 & -9.657 $\pm$ 0.015 & 4.835 $\pm$ 0.011 & 35 & 15.88 $\pm$ 1.13 & 17.68 & 251.02 & 7.98 & 406$^{+58}_{-38}$ \\
NGC 2527 & 121.265 & -28.107 & 8.919 & -0.1 & 0.195 & 572.01 & 0.263 $\pm$ 0.012 & 0.475 $\pm$ 0.073 & 368 & 1.562 $\pm$ 0.003 & -5.543 $\pm$ 0.01 & 7.319 $\pm$ 0.009 & 39 & 39.53 $\pm$ 0.97 & 25.91 & 343.01 & 9.3837 & 6383$^{+3984}_{-2286}$ \\
NGC 2546 & 123.082 & -37.661 & 7.962 & 0.01 & 0.464 & 920.03 & 0.297 $\pm$ 0.075 & 0.221 $\pm$ 0.081 & 445 & 1.067 $\pm$ 0.002 & -3.709 $\pm$ 0.006 & 3.942 $\pm$ 0.006 & 6 & 12.63 $\pm$ 0.48 & 19.99 & 554.46 & 10.69 & 1053$^{+195}_{-126}$ \\
NGC 3228 & 155.378 & -51.814 & 7.424 & 0.01 & 0.131 & 473.15 & 0.183 $\pm$ 0.029 & 0.257 $\pm$ 0.055 & 145 & 2.068 $\pm$ 0.007 & -14.856 $\pm$ 0.019 & -0.609 $\pm$ 0.017 & 16 & 16.96 $\pm$ 2.18 & 19.91 & 120.74 & 6.39 & 194$^{+28}_{-19}$ \\
NGC 5822 & 226.051 & -54.366 & 9.02 & 0.08 & 0.339 & 843.33 & 0.045 $\pm$ 0.03 & 0.316 $\pm$ 0.098 & 620 & 1.206 $\pm$ 0.001 & -7.474 $\pm$ 0.007 & -5.492 $\pm$ 0.007 & 54 & -26.16 $\pm$ 1.44 & 23.11 & 748.4 & 12.6 & 10244$^{+6032}_{-3493}$ \\
NGC 6475 & 268.447 & -34.841 & 8.477 & 0.07 & 0.152 & 279.77 & 0.118 $\pm$ 0.011 & 0.21 $\pm$ 0.048 & 980 & 3.594 $\pm$ 0.002 & 3.09 $\pm$ 0.016 & -5.362 $\pm$ 0.015 & 130 & -14.4 $\pm$ 1.24 & 32.42 & 845.29 & 13.26 & 2881$^{+992}_{-614}$\\
NGC 6793 & 290.817 & 22.159 & 8.82 & 0.0 & 0.95 & 536.54 & 0.264 $\pm$ 0.015 & 0.286 $\pm$ 0.141 & 208 & 1.674 $\pm$ 0.003 & 3.775 $\pm$ 0.011 & 3.537 $\pm$ 0.011 & 17 & -17.49 $\pm$ 2.25 & 18.39 & 192.06 & 8.03 & 4204$^{+2694}_{-1539}$\\
NGC 6940 & 308.626 & 28.728 & 9.02 & 0.15 & 0.45 & 1041.36 & 0.203 $\pm$ 0.009 & 0.455 $\pm$ 0.11 & 713 & 0.963 $\pm$ 0.002 & -1.952 $\pm$ 0.005 & -9.434 $\pm$ 0.006 & 39 & 8.81 $\pm$ 0.78 & 51.16 & 780.13 & 12.34 & 10346$^{+6051}_{-3507}$ \\
NGC 7243 & 333.788 & 49.83 & 8.022 & 0.03 & 0.573 & 888.38 & 0.239 $\pm$ 0.043 & 0.215 $\pm$ 0.118 & 401 & 1.128 $\pm$ 0.002 & 0.438 $\pm$ 0.006 & -2.883 $\pm$ 0.007 & 3 & 63.94 $\pm$ 0.09 & 19.57 & 456.74 & 10.59 & 981$^{+220}_{-141}$ \\
Pleiades & 56.601 & 24.114 & 7.937 & 0.0 & 0.14 & 135.96 & 0.262 $\pm$ 0.052 & 0.225 $\pm$ 0.079 & 1410 & 7.377 $\pm$ 0.006 & 19.96 $\pm$ 0.034 & -45.429 $\pm$ 0.042 & 262 & 5.66 $\pm$ 0.98 & 58.78 & 687.99 & 12.36 & 1211$^{+193}_{-126}$\\
Roslund 6 & 307.185 & 39.798 & 7.976 & 0.0 & 0.144 & 335.43 & 0.42 $\pm$ 0.056 & 0.324 $\pm$ 0.115 & 230 & 2.832 $\pm$ 0.003 & 5.98 $\pm$ 0.016 & 2.143 $\pm$ 0.013 & 39 & -6.74 $\pm$ 2.65 & 22.19 & 181.42 & 7.15 & 508$^{+156}_{-97}$ \\
Ruprecht 98 & 179.715 & -64.581 & 8.756 & 0.0 & 0.5 & 465.37 & 0.471 $\pm$ 0.02 & 0.592 $\pm$ 0.06 & 132 & 2.081 $\pm$ 0.005 & -4.158 $\pm$ 0.044 & -8.564 $\pm$ 0.029 & 6 & -2.27 $\pm$ 1.83 & 24.77 & 142.2 & 6.2 & 3256$^{+2102}_{-1199}$ \\
Ruprecht 161 & 152.4 & -61.259 & 8.35 & 0.0 & 0.344 & 904.9 & 0.439 $\pm$ 0.009 & 0.166 $\pm$ 0.191 & 90 & 1.096 $\pm$ 0.004 & -10.322 $\pm$ 0.018 & 3.827 $\pm$ 0.019 & 3 & 19.73 $\pm$ 1.53 & 23.49 & 98.45 & 5.53 & 891$^{+484}_{-284}$ \\
Stock 1 & 294.146 & 25.163 & 8.676 & 0.0 & 0.387 & 386.37 & 0.344 $\pm$ 0.017 & 0.419 $\pm$ 0.1 & 132 & 2.455 $\pm$ 0.003 & 6.019 $\pm$ 0.021 & 0.24 $\pm$ 0.021 & 24 & -18.73 $\pm$ 4 & 26.93 & 141.18 & 6.86 & 2527$^{+1574}_{-903}$\\
Stock 12 & 353.923 & 52.685 & 8.3 & 0.0 & 0.27 & 416.48 & 0.375 $\pm$ 0.02 & 0.355 $\pm$ 0.14 & 147 & 2.295 $\pm$ 0.004 & 8.554 $\pm$ 0.015 & -1.925 $\pm$ 0.016 & 19 & -1.57 $\pm$ 2.82 & 20.12 & 129.78 & 6.8 & 854$^{+420}_{-249}$ \\
Teutsch 35 & 294.091 & 35.742 & 8.38 & 0.15 & 0.0 & 479.95 & 0.289 $\pm$ 0.03 & 0.428 $\pm$ 0.085 & 172 & 2.006 $\pm$ 0.003 & -0.632 $\pm$ 0.014 & -2.471 $\pm$ 0.015 & 18 & -10.58 $\pm$ 2.59 & 33.17 & 150.83 & 6.54 & 1102$^{+561}_{-332}$\\
Trumpler 2 & 39.232 & 55.905 & 8.023 & 0.0 & 0.954 & 703.07 & 0.161 $\pm$ 0.024 & 0.193 $\pm$ 0.077 & 253 & 1.452 $\pm$ 0.003 & 1.523 $\pm$ 0.009 & -5.429 $\pm$ 0.012 & 13 & -1.21 $\pm$ 1.4 & 18.14 & 309.46 & 8.8 & 758$^{+201}_{-127}$\\
Trumpler 10 & 131.943 & -42.566 & 7.74 & -0.12 & 0.138 & 365.76 & 0.352 $\pm$ 0.021 & 0.298 $\pm$ 0.079 & 1101 & 2.325 $\pm$ 0.003 & -12.511 $\pm$ 0.017 & 6.613 $\pm$ 0.009 & 81 & 21.87 $\pm$ 0.54 & 24.24 & 734.08 & 11.06 & 1094$^{+115}_{-76}$ \\
            \hline
        \end{tabular}
        }
    \end{landscape}
}

The 46 open clusters studied in this work are highlighted in black in Figure~\ref{fig:clus_sel} [bottom]. {Their identified members and colour magnitude diagrams (CMDs) in \textit{Gaia} filters are presented in Appendix~\ref{sect:members}.} Their cluster parameters are tabulated in Table~\ref{table:clus}, including the cluster age, extinction, distance and metallicity taken from \citet{Bossini19} and the ellipticities from \citet{Hu21}. We also note that the fractional uncertainty in parallax for our cluster members varies with distance with typical values of 0.008, 0.021, 0.066 \& 0.055 for IC 2602, NGC 6475, NGC 6940 \& Gulliver 44 respectively which are at respective distances of 152.34~pc, 279.77~pc, 1041.36~pc \& 1258.92~pc. The radial physical extent (R$\rm_{C}$) of the identified cluster is noted as the physical distance of the furthest identified member from the cluster center. Figure~\ref{fig:gal} shows the spatial distribution of the identified members of these 46 clusters in Galactic coordinates.

\section{Analysis}
\label{sect:analysis}

Each of the 46 clusters are searched for tidal features. Using Teutsch~35 as an example in the following subsections, we demonstrate the steps applied to each of the clusters to search for tidal features.  

\subsection{Color Magnitude Diagram}
\label{sect:cmd}

\begin{figure}
	\includegraphics[width=\columnwidth]{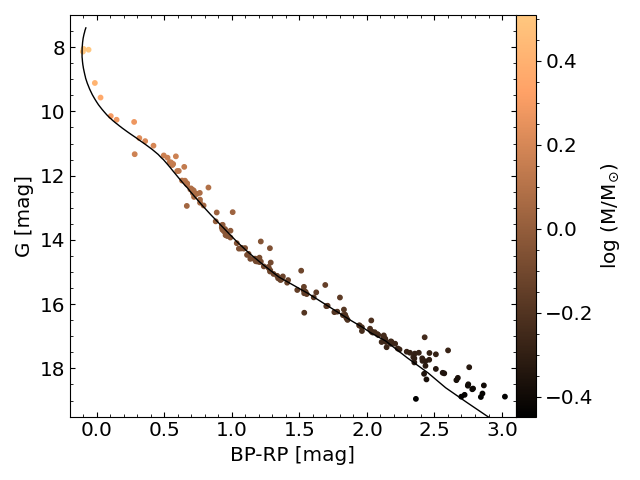}
    \caption{CMD of all the identified members in Teutsch~35, coloured by their stellar masses estimated from the PARSEC isochrone (black).}
    \label{fig:cmd}
\end{figure}

The CMD of the identified members of Teutsch~35 is shown in Figure~\ref{fig:cmd}. The PARSEC stellar evolution isochrone\footnote{\url{http://stev.oapd.inaf.it/cgi-bin/cmd}} \citep{Bressan12,Chen14} for the \textit{Gaia} filters has been plotted for the cluster age, extinction, distance and metallicity (see Table~\ref{table:clus}). Stellar mass of each star was estimated from the PARSEC isochrone and is marked in Figure~\ref{fig:cmd}. {The total observed mass of the cluster members (hereafter, photometric mass) is m$\rm_{p}=150.83$~M$_{\sun}$.} 

\subsection{Tidal radius}
\label{sect:tr}

\begin{figure}
	\includegraphics[width=0.9\columnwidth]{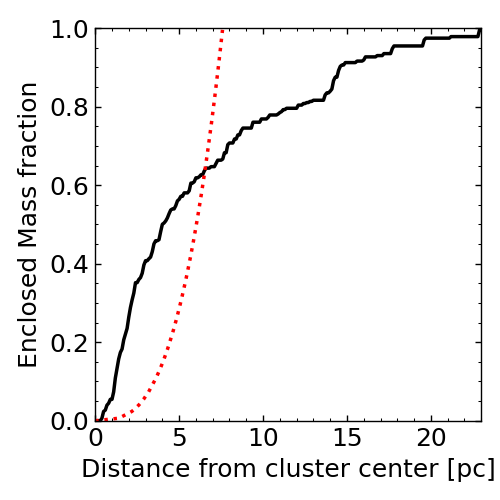}
    \caption{Radial enclosed mass profile (black solid) and tidal radius profile (red dotted) for corresponding enclosed mass for Teutsch~35. The intersection of these profiles identifies the tidal radius of the cluster.}
    \label{fig:tidal_radius}
\end{figure}

\begin{figure}
	\includegraphics[width=\columnwidth]{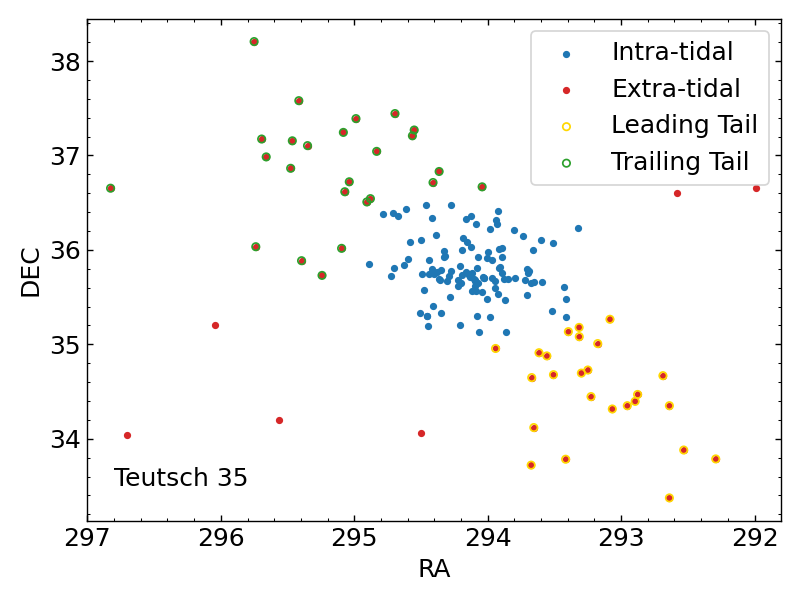}
    \caption{Spatial distribution of the identified members in Teutsch~35 with the intra-tidal (blue), extra-tidal (red), leading tail (encircled in yellow) and trailing tail (encircled in green) members marked.}
    \label{fig:spat_t35}
\end{figure}

\begin{figure*}
	\includegraphics[width=\textwidth]{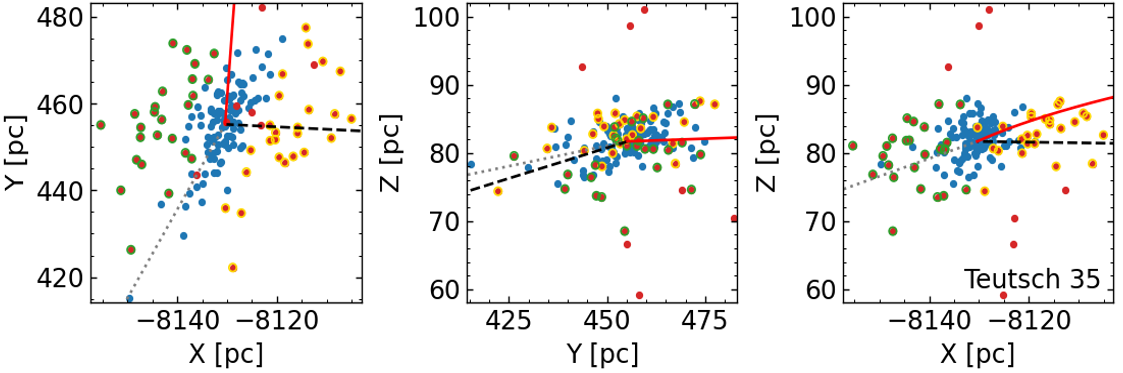}
    \caption{Tri-dimensional projections in the galactocentric coordinate system of the identified members in Teutsch~35 with the sub-regions as marked in Figure~\ref{fig:spat_t35}. The directions of the cluster orbit (red solid line), to the Galactic center (black dashed line) and to the line-of-sight (grey dotted) are shown.}
    \label{fig:orbit}
\end{figure*}

While extended tails are clearly visible in the spatial distribution of Teutsch~35 (Figure~\ref{fig:mlmoc} [right]), we need to estimate the tidal radius (r$\rm_{t}$) of the cluster in order to ascertain the extra-tidal nature of these tails. Estimating the tidal radius from fitting the King function \citep{King66} to the radial density of the cluster is not appropriate for a dissolving cluster with extended structures as such a fit would overestimate r$\rm_{t}$ (See Appendix~\ref{sect:tarricq}). It is because the King model is appropriate only for a system in equilibrium, inappropriate for extended systems such as open clusters with tidal tails, though still appropriate within r$\rm_{t}$ for older open clusters in equilibrium \citep{Elson87}. Thus, r$\rm_{t}$ may be calculated from the enclosed mass within r$\rm_{t}$ ignoring the mass contribution from beyond \citep{Meingast21}. We thus calculate r$\rm_{t}$ following \citet{Pinfield98} as:
\begin{equation}
    $$\rm r_{t}^{3} = \rm \frac{G M_{enc}}{2(A-B)^{2}}$$
\end{equation}
where G is the gravitational constant, A and B are the
Oort constants (A~$=15.3\pm0.4$ km/s/kpc, B~$=-11.9\pm0.4$ km/s/kpc; \citealt{Bovy17}). Following \citet{Meingast21}, we obtain r$\rm_{t}$ by comparing the radial fractional mass distribution of the cluster members (black solid line in Figure~\ref{fig:tidal_radius}) to that of the r$\rm_{t}$ corresponding to the enclosed mass (red dotted line in Figure~\ref{fig:tidal_radius}). Where the two curves meet in Figure~\ref{fig:tidal_radius} is the r$\rm_{t}$ of the cluster. Note that we have assumed the observed photometric mass to be the enclosed mass but an error of $\sim$3\% would need to be added to account for unobserved stars\footnote{We saw in \citet{Bhattacharya21} that for NGC~752, by extrapolating the fitted mass function to the Hydrogen burning limit, $\sim$3\% of the stellar mass was unobserved. A similar value is expected for the clusters in this work though it may be slightly larger for more distant clusters.}. Including the error on the Oort constants, the uncertainty on r$\rm_{t}$ would be $\sim$5\%.  We calculate r$\rm_{t}=0.7807^{\circ}=6.54$~pc at the distance of Teutsch~35. The identified members within and beyond r$\rm_{t}$ {in the 2D spatial projection} are classified as intra-tidal and extra-tidal members respectively  (Figure~\ref{fig:spat_t35}). The extra-tidal members go out to 33.17~pc at the distance of the cluster. The extended features beyond the tidal radius are clearly visible in Figure~\ref{fig:spat_t35} but {to check their alignment with respect to the cluster orbit, we explore their positions in galactocentric coordinates.}

\begin{table*}
\caption{Mean galactocentric positions (columns 2--4) and space velocities (columns 5--7) for the members identified in each of the 20 open clusters having tidal tails. Columns 8--9 note the extent of the tidal tails.}
\centering
\adjustbox{max width=\textwidth}{
\begin{tabular}{ccccccccc}
  \hline
    Name & X & Y & Z & U & V & W & $\rm L_{lead}$ & $\rm L_{lag}$ \\
     & kpc & kpc & kpc & km/s & km/s & km/s & pc & pc\\    
  \hline
Alessi 2 & -8835.2 & 280.22 & 89.85 & 20.92 & 240.72 & 2.73 & 16.01 & 16.37 \\
Alessi 24 & -7899.21 & -235.89 & -100.44 & 11.3 & 225.5 & -4.69 & 31.31 & 20.17\\
Alessi 6 & -7709.22 & -625.82 & -65.22 & -31.07 & 221.18 & 5.67 & 15.02 & 19.11 \\
ASCC 101 & -8155.6 & 356.7 & 99.15 & 3.63 & 230.32 & 3.37 & 11.09 & 9.15 \\
BH 164 & -8012.24 & -295.46 & -24.91 & -2.93 & 229.69 & -4.84 & 13.78 & 15.37 \\
IC 2602 & -8250.29 & -142.21 & 7.35 & 4.17 & 225.74 & 6.99 & 23.54 & 14.03 \\
M 39 & -8310.29 & 296.43 & 9.35 & 39.85 & 239.98 & -5.26 & 27.09 & 27.64 \\
NGC 1662 & -8672.93 & -51.36 & -122.71 & 25.36 & 245.49 & 8.19 & 20.47 & 18.34 \\
NGC 2451B & -8408.19 & -338.15 & -22.64 & -5.95 & 236.39 & -4.22 & 17.1 & 17.68 \\
NGC 2527 & -8554.61 & -573.33 & 42.67 & -27.86 & 220.45 & 6.96 & 25.91 & 25.02 \\
NGC 3228 & -8211.02 & -466.18 & 58.17 & -11.65 & 224.88 & -10.61 & 19.92 & 16.88 \\
NGC 5822 & -7664.52 & -504.01 & 70.38 & -29.28 & 234.0 & 2.02 & 22.89 & 23.11 \\
NGC 6475 & -8023.37 & -20.11 & -1.84 & -2.3 & 242.59 & 1.82 & 32.42 & 29.2 \\
NGC 6940 & -7950.89 & 952.02 & -107.15 & 54.77 & 236.97 & -12.74 & 49.17 & 41.7 \\
NGC 7243 & -8432.27 & 849.45 & -63.06 & 8.41 & 308.31 & -8.19 & 19.57 & 18.11 \\
Pleiades & -8420.88 & 29.22 & -32.62 & 6.31 & 217.21 & -6.28 & 41.19 & 34.5 \\
Roslund 6 & -8230.6 & 343.91 & 24.42 & 3.07 & 240.76 & 1.72 & 22.19 & 18.28 \\
Ruprecht 98 & -8081.86 & -422.02 & 2.64 & 7.76 & 244.51 & -12.71 & 24.03 & 19.48 \\
Stock 12 & -8455.35 & 394.2 & -42.45 & -0.36 & 237.33 & -0.78 & 17.62 & 20.12 \\
Teutsch 35 & -8130.39 & 455.2 & 81.71 & 14.83 & 233.89 & 4.86 & 28.11 & 30.31 \\
  \hline
\end{tabular}
\label{table:orbit}
}
\end{table*}

\subsection{Cluster Orbit}
\label{sect:orbit}

Similar to \citet{Tang19} for Coma Berenices and \citet{Bhattacharya21} for NGC~752, we explore the position of cluster members of Teutsch~35 with respect to the cluster orbit in a galactocentric coordinate system. These coordinates have a positive x direction pointing from the position of the Sun  towards the {GC}; the y-axis points toward \textit{l} = 90$^{\circ}$ and the z-axis roughly points toward \textit{b} = 90$^{\circ}$.  We take the distance of the Sun to the GC as 8.3 kpc, and solar motion parameters as U$_{\odot}=12.9$~km~s$^{-1}$, V$_{\odot}=245.6$~km~s$^{-1}$ and W$_{\odot}=7.78$~km~s$^{-1}$. \citep{Reid04}.  A known effect of distance estimation from parallax measurements (required for conversion to the galactocentric coordinate system) is that the morphology of star clusters appears to be stretched along the line of sight when distances are obtained by simple parallax inversion \citep[E.g.][]{Pang21}. This effect becomes more prominent with increasing cluster distance. To mitigate this issue, we use the photogeometric distances to our cluster members for \textit{Gaia} EDR3 as found by \citet{Bj21}.  They estimate stellar distances following a probabilistic approach using a prior constructed from a three-dimensional model of the Galaxy, taking into account interstellar extinction, the \textit{Gaia} variable magnitude limit, and the color and apparent magnitude of each star.

Figure~\ref{fig:orbit} shows the X-Y (left), Y-Z (middle) and X-Z (right) projections of the cluster members. We compute the average position of all member stars as the cluster mean position and its mean space velocity for only those sources having RV measurements (see Table~\ref{table:orbit}). These values are utilised to calculate the cluster orbit using the python package galpy \citep{Bovy15} which accounts for solar motion and includes the Milky Way potential model \textit{MWPotential2014} (a fitted Galactic potential model incorporating the bulge, disc and halo). Orbits are integrated 20 Myr forward in time. 

Figure~\ref{fig:orbit} shows the orbit direction (red solid line) of Teutsch~35 in the three galactocentric projections. Also shown are the line-of-sight direction (grey dotted line) and the direction towards the GC (black dashed line). {Clearly the cluster still appears extended along the line-of-sight, including the extra-tidal members, most prominent in the X-Y plane in Figure~\ref{fig:orbit}. Despite the extension along the line-of-sight, the  extra-tidal members encircled in yellow in Figure~\ref{fig:orbit} point towards the GC in the X-Y plane and intermediate between the direction of the cluster orbit and the GC in the X-Z plane.} These form the leading tail of Teutsch~35 while those encircled in green point in the opposite direction and form its trailing tail. As the cluster orbit direction is nearly opposite to that of the GC in the Y-Z plane, the tidal tail members do not occupy distinct areas in this plane. Thus, we find that Teutsch~35 is a  dissolving open cluster exhibiting tidal features. Figure~\ref{fig:spat_t35} shows the distinct sub-regions of Teutsch~35, its intra- and extra-tidal regions as well as the leading and trailing tails. The leading and trailing tails extend to 28.11~pc and 30.31~pc respectively at the distance of the cluster. {We note that to confirm if these tails are indeed tidally formed, their kinematic dissipation along the cluster orbit needs to be checked. Such a check is only possible when many of the tail members have 6D astrometric information such as the nearby Coma~Berenices studied by \citet{Tang19} but not possible for Teutsch~35 or most of the other clusters studied in this work.}

\begin{figure}
	\includegraphics[width=\columnwidth]{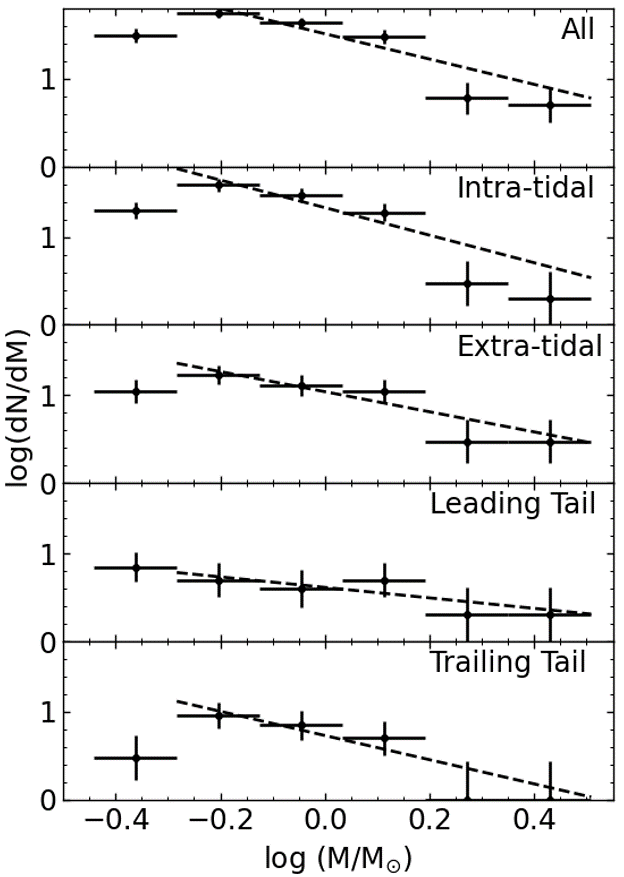}
    \caption{The mass function derived from the PARSEC iscochrones for the entire Teutsch~35 (all) and other sub-regions (as marked {in Figure~\ref{fig:spat_t35}}). The dashed black line shows the fitted mass function where the \textit{Gaia} EDR3 data is almost complete. The Poissonian uncertainties have been marked.}
    \label{fig:mf}
\end{figure}

\subsection{Mass function}
\label{sect:mf}

The mass function of Teutsch~35 is fitted above the stellar mass value 0.5~M$_{\sun}$, which corresponds to the turn-off point, above which value the Salpeter and Kroupa initial mass functions agree \citep{Kroupa01}. For Teutsch~35, this corresponds to G~=~17.5~mag, much brighter than the 90\% completeness limit (see Appendix~\ref{sect:comp}). 
The mass function of all the identified members of the cluster is shown in Figure~\ref{fig:mf} [top]. Using the relation log(dN/dM)~=~$-$~(1~+~$\chi$)~$\times$~log(M)~+~constant, where dN represents the number of stars in a mass bin dM with central mass M, we obtain the slope of the mass function for Teutsch~35, $\chi=0.44\pm0.38$. Such a low value of $\chi$, as compared to its value of $\chi=1.37$ as derived by \citet{salpeter55} for the solar neighbourhood, has previously been considered as an indication of a dissolving cluster that has undergone mass segregation \citep[E.g.][]{Bhattacharya19}. We further obtain the mass functions for the four sub-regions of Teutsch~35 (intra-tidal, extra-tidal, leading tail and trailing tail) as shown in Figure~\ref{fig:mf}.  The fitted slopes of the mass functions are $\chi=0.57\pm0.46$, $0.14\pm0.29$, $-0.4\pm0.28$ and $0.37\pm0.37$ for the intra-tidal, extra-tidal, leading tail and trailing tail respectively. {Since the stripped cluster members are expected to span an even larger spatial region \citep[E.g.][]{Jerabkova21}, the mass function estimates of the extra-tidal and tail regions are highly uncertain.}

\subsection{Tidal tails in searched open clusters}
\label{sect:tt}

\begin{figure*}
	\includegraphics[width=\textwidth]{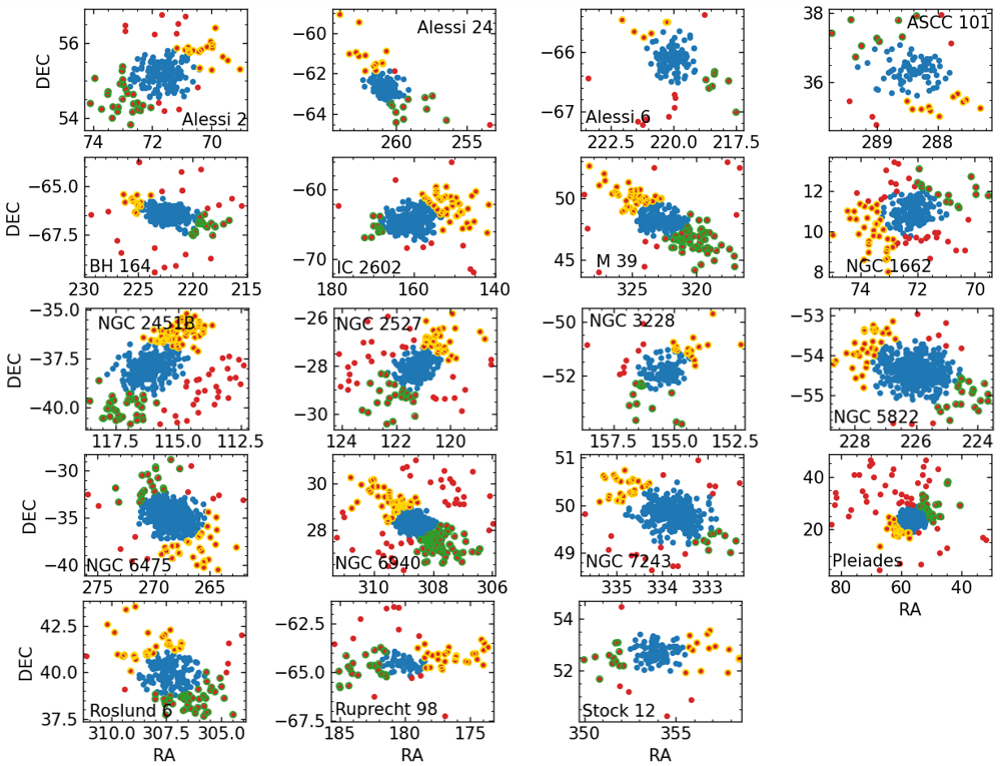}
    \caption{Spatial distribution of the identified members for the other 19 open clusters with tidal tails. The sub-regions are as marked in Figure~\ref{fig:spat_t35}.}
    \label{fig:tt}
\end{figure*}

The m$\rm_{p}$ and r$\rm_{t}$ values of the other 45 open clusters were obtained in the same way as for Teutsch~35 in Section~\ref{sect:tr} and have been tabulated in Table~\ref{table:clus}. The Pleiades has the largest physical coronal radius (R$\rm_{C}=58.78$~pc) while Alessi~Teutsch~3 has the smallest (R$\rm_{C}=9.88$~pc). From the spatial distribution of cluster members, extended features beyond the tidal radius are visually identified in each cluster and tentatively marked as tidal tail candidates. {As in Section~\ref{sect:orbit} for Teutsch~35, we check the alignment of the cluster members with the cluster orbits in galactocentric coordinates. 19 clusters (see Figure~\ref{fig:orbit_append}) meet the tidal tail criteria. The alignment of their tails with the cluster orbit is clear despite the line-of-sight elongation for most of these clusters.} Note that we are not sensitive to the presence of tidal tails along the line-of-sight given this apparent elongation. These 19 clusters, in addition to Teutsch~35, have thus been identified to have extended tidal tails and their spatial positions are shown in Figure~\ref{fig:tt}. The members of these clusters not belonging to either of the tails but present beyond the tidal radius form part of the cluster corona. Table~\ref{table:orbit} notes their mean galactocentric positions and space velocities as well as the physical extent of their tidal tails. NGC~6940 has the largest physical tidal tail length (L$\rm_{lead}=49.17$~pc; L$\rm_{lag}=41.7$~pc) while ASCC~101 has the smallest (L$\rm_{lead}=11.09$~pc; L$\rm_{lag}=9.15$~pc).

\begin{figure*}
	\includegraphics[width=0.95\textwidth]{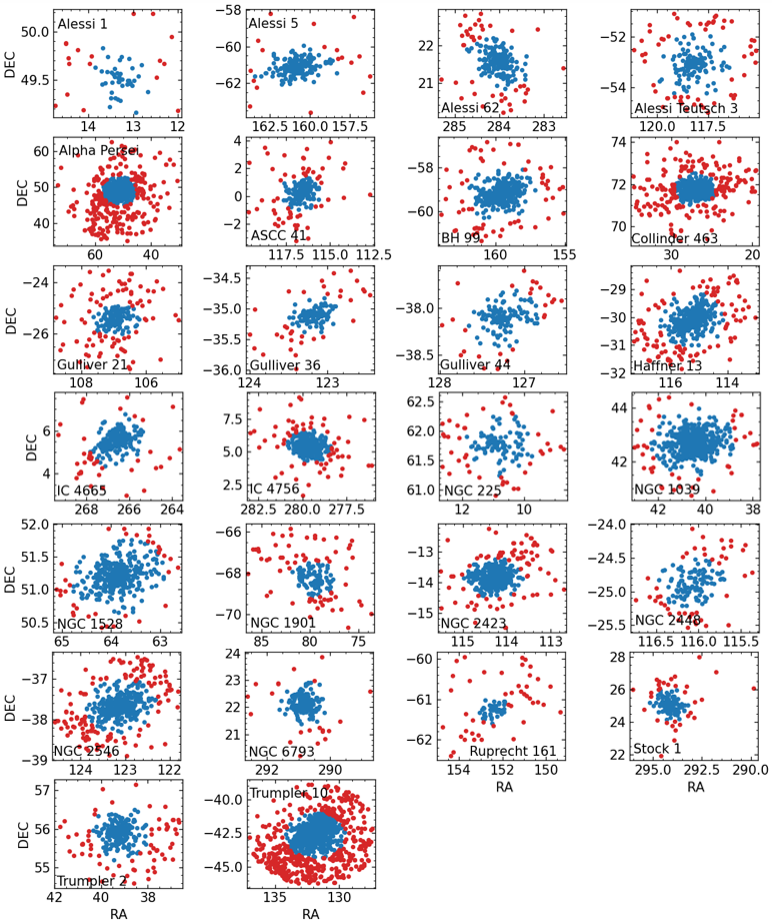}
    \caption{Spatial distribution of the identified members for the {26} open clusters which have a corona but no clear tidal tails. The intra-tidal (blue) and extra-tidal (red) members are marked.}
    \label{fig:corona}
\end{figure*}

The majority of the other 26 open clusters only have a corona of extra-tidal members, without any clear tidal extension. Some of these have seemingly extended features observed spatially (namely, Alessi~62, ASCC~41, Gulliver~36, Gulliver~44, NGC~1528, NGC~2448 and Ruprecht~161) but these features do not align securely with the cluster orbit or GC direction and are thus not classified as having tidal tails. The spatial positions of these 25 open clusters along with their extra tidal members are shown in Figure~\ref{fig:corona}. 

\section{Discussion}
\label{sect:disc}

{We reiterate that members identified by ML-MOC for each cluster would by design have gaussian distributions in their proper motion and parallax values (Section~\ref{sect:data}). However, it is already known that tail members identified at larger distances from the center of any cluster would necessarily span separate regions in the proper motion and parallax parameter spaces depending on the cluster properties \citep[E.g.][]{Jerabkova21}. So tail members beyond a certain radius would not be identified by ML-MOC and this limit would be distinct for each cluster depending on its age, distance and other properties. \citet{Boffin22} explored the distribution of a synthetic cluster model akin to NGC~752. They find that ML-MOC identifies only those tail members in NGC~752 \citep{Bhattacharya21} that are very close to the cluster core, though those identified are consistent with the modelled tail region. Membership determination algorithms constrained by tailored synthetic models made for NGC~752 are required to identify the full extent of its tidal tails \citep{Boffin22}. Such synthetic models are beyond the scope of this paper and our goal is to reliably identify clusters with tidal tails using ML-MOC, knowing that the length of tidal tails found in this work are a lower limit on the actual cluster tidal tail length. We discuss our results keeping this in mind.}

\begin{table*}
\caption{Mass function slope and mass segregation class for the 46 open clusters studied in this work. Column 1: Name; Column 2--4: Mass function slope for all cluster members and those within and outside the tidal radius; Columns 5--6: Mass function slopes for the tails of the open clusters where determinable (errors are not quoted if indeterminate); Column 7: Mass segregation class assigned based on the $\rm\lambda_{MSR}$ vs N$\rm_{massive}$ profile shapes.}
\label{table:seg}
    \centering 
    \adjustbox{max width=\textwidth}{
    \begin{tabular}{ccccccc}
    \hline
Name & $\rm\chi_{all}$ & $\rm\chi_{intra}$ & $\rm\chi_{extra}$ & $\rm\chi_{lead}$ & $\rm\chi_{lag}$ & Class\\
\hline
Alessi 1 & -1.0 $\pm$ 0.17 & -1.36 $\pm$ 0.16 & -0.08 $\pm$ 0.34 & -- & -- & 3\\
Alessi 2 & 0.48 $\pm$ 0.26 & 0.41 $\pm$ 0.17 & 1.01 $\pm$ 0.32 & 0.95 $\pm$ 0.7 & 1.55 $\pm$ 0.25 & 3\\
Alessi 24 & 0.85 $\pm$ 0.14 & 0.68 $\pm$ 0.19 & 0.95 $\pm$ 0.13 & -- & -- & 3 \\
Alessi 5 & 0.56 $\pm$ 0.24 & 0.45 $\pm$ 0.25 & 0.74 $\pm$ 0.25 & -- & -- & 3 \\
Alessi 6 & -0.59 $\pm$ 0.3 & -0.5 $\pm$ 0.3 & 0.88 $\pm$ 0.49 & 1.34  & 0.37 $\pm$ 0.23 & 4\\
Alessi 62 & 0.14 $\pm$ 0.29 & -0.12 $\pm$ 0.31 & 1.19 $\pm$ 0.29 & -- & -- & 2 \\
Alessi Teutsch 3 & 0.91 $\pm$ 0.34 & 0.48 $\pm$ 0.25 & 0.96 $\pm$ 0.28 & -- & -- & 3 \\
Alpha Persei & 0.38 $\pm$ 0.12 & 0.33 $\pm$ 0.07 & 0.48 $\pm$ 0.31 & -- & -- & 2 \\
ASCC 101 & -0.11 $\pm$ 0.3 & -0.16 $\pm$ 0.23 & -0.04 $\pm$ 0.88 & 1.26  & -1.25 $\pm$ 0.37 & 3\\
ASCC 41 & 0.16 $\pm$ 0.05 & 0.16 $\pm$ 0.23 & 0.61 $\pm$ 0.42 & -- & -- & 3 \\
BH 164 & 0.54 $\pm$ 0.1 & 0.49 $\pm$ 0.09 & 0.71 $\pm$ 0.14 & 0.81 $\pm$ 0.14  & 0.2 $\pm$ 0.52  & 4\\
BH 99 & 0.46 $\pm$ 0.26 & 0.49 $\pm$ 0.18 & 1.0 $\pm$ 0.11 & -- & -- & 3 \\
Collinder 463 & 0.11 $\pm$ 0.16 & -0.04 $\pm$ 0.12 & 0.57 $\pm$ 0.2 & -- & -- & 2 \\
Gulliver 21 & 0.34 $\pm$ 0.13 & 0.09 $\pm$ 0.14 & 0.93 $\pm$ 0.26 & -- & -- & 3 \\
Gulliver 36 & 0.28 $\pm$ 0.26 & -0.01 $\pm$ 0.41 & 1.81 $\pm$ 0.29 & -- & -- & 2 \\
Gulliver 44 & 0.41 $\pm$ 0.44 & 0.41 $\pm$ 0.54 & 0.28 $\pm$ 0.48 & -- & -- & 3 \\
Haffner 13 & 0.51 $\pm$ 0.17 & 0.57 $\pm$ 0.18 & 0.25 $\pm$ 0.47 & -- & -- & 3 \\
IC 2602 & 0.26 $\pm$ 0.1 & 0.23 $\pm$ 0.11 & 0.31 $\pm$ 0.19 & -- & -- & 2 \\
IC 4665 & 0.32 $\pm$ 0.03 & 0.27 $\pm$ 0.05 & 0.69 $\pm$ 0.1 & -- & -- & 3 \\
IC 4756 & -0.7 $\pm$ 0.23 & 0.22 $\pm$ 0.33 & 0.6 $\pm$ 0.46 & -- & -- & 1 \\
M39 & -0.07 $\pm$ 0.12 & -0.12 $\pm$ 0.08 & -0.28 $\pm$ 0.79 & 0.35 $\pm$ 0.79 & 0.41 $\pm$ 0.79 & 2\\
NGC 225 & 0.74 $\pm$ 0.2 & 0.3 $\pm$ 0.17 & 1.64 $\pm$ 0.15 & -- & -- & 2 \\
NGC 1039 & 0.56 $\pm$ 0.22 & 0.58 $\pm$ 0.27 & 1.38 $\pm$ 0.31 & -- & -- & 2 \\
NGC 1528 & 0.32 $\pm$ 0.11 & 0.26 $\pm$ 0.19 & 0.36 $\pm$ 0.62 & -- & -- & 1 \\
NGC 1662 & 0.19 $\pm$ 0.21 & 0.04 $\pm$ 0.13 & 0.67 $\pm$ 0.69 & 0.63 $\pm$ 0.33 & 1.47 $\pm$ 0.46 & 3\\
NGC 1901 & -0.44 $\pm$ 0.08 & -1.08 $\pm$ 0.14 & 1.08 $\pm$ 0.2 & -- & -- & 2 \\
NGC 2423 & -0.37 $\pm$ 0.05 & -0.55 $\pm$ 0.16 & 0.99 $\pm$ 0.35 & -- & -- & 2 \\
NGC 2448 & 0.22 $\pm$ 0.33 & 0.03 $\pm$ 0.27 & 0.67 $\pm$ 0.69 & -- & -- & 3 \\
NGC 2451B & 0.57 $\pm$ 0.33 & 0.54 $\pm$ 0.38 & 0.62 $\pm$ 0.24 & 0.28 $\pm$ 0.31 & 1.24 $\pm$ 0.31  & 4\\
NGC 2527 & -0.27 $\pm$ 0.16 & -0.52 $\pm$ 0.16 & 1.01 $\pm$ 0.58 & 0.75 $\pm$ 0.27 & 1.18 $\pm$ 0.27 & 2\\
NGC 2546 & 0.12 $\pm$ 0.16 & 0.07 $\pm$ 0.17 & 0.33 $\pm$ 0.34 & -- & -- & 2 \\
NGC 3228 & -0.06 $\pm$ 0.31 & -0.07 $\pm$ 0.29 & -0.03 $\pm$ 0.47 & 0.14 $\pm$ 0.24 & -0.53 $\pm$ 0.7 & 4 \\
NGC 5822 & -0.81 $\pm$ 0.29 & -0.84 $\pm$ 0.21 & 0.26 $\pm$ 0.71 & 0.13 $\pm$ 1.64 & 0.17 $\pm$ 1.35 & 1\\
NGC 6475 & 0.26 $\pm$ 0.25 & 0.22 $\pm$ 0.32 & 0.99$\pm$ 0.33 & 1.01 $\pm$ 0.9 & 0.08 $\pm$ 0.37 & 1\\
NGC 6793 & 0.4 $\pm$ 0.17 & 0.28 $\pm$ 0.17 & 1.9 $\pm$ 0.1 & -- & -- & 2 \\
NGC 6940 & -0.04 $\pm$ 0.06 & -0.22 $\pm$ 0.08 & 0.65 $\pm$ 0.43 & 0.03 $\pm$ 0.62 & 1.63 $\pm$ 1.04 & 2\\
NGC 7243 & 0.78 $\pm$ 0.14 & 0.73 $\pm$ 0.16 & 1.23 $\pm$ 0.32 & 0.44  & 1.49 $\pm$ 0.54 & 3\\
Pleiades & 0.69 $\pm$ 0.18 & 0.6 $\pm$ 0.22 & 1.07 $\pm$ 0.11 & -- & -- & 3 \\
Roslund 6 & 0.14 $\pm$ 0.2 & -0.08 $\pm$ 0.25 & 0.38 $\pm$ 0.24 & 0.2 $\pm$ 0.33 & 0.41 $\pm$ 0.26 & 1\\
Ruprecht 98 & -0.68 $\pm$ 0.4 & -1.15 $\pm$ 0.45 & -0.03 $\pm$ 0.28 & -0.35 $\pm$ 0.38 & 0.21 $\pm$ 0.38 & 2\\
Ruprecht 161 & 0.49 $\pm$ 0.18 & -0.02 $\pm$ 0.33 & 0.95 $\pm$ 0.2 & -- & -- & 3 \\
Stock 1 & -0.6 $\pm$ 0.06 & -0.68 $\pm$ 0.08 & -0.24 $\pm$ 0.1 & -- & -- & 1 \\
Stock 12 & 0.11 $\pm$ 0.18 & 0.03 $\pm$ 0.1 & -0.15 $\pm$ 1.14 & 0.43  & -- & 2\\
Teutsch 35 & 0.44 $\pm$ 0.38 & 0.57 $\pm$ 0.46 & 0.14 $\pm$ 0.3 & -0.4 $\pm$ 0.28 & 0.38 $\pm$ 0.37 & 4\\
Trumpler 2 & 0.25 $\pm$ 0.18 & 0.19 $\pm$ 0.24 & 0.56 $\pm$ 0.2 & -- & -- & 3 \\
Trumpler 10 & 0.24 $\pm$ 0.25 & 0.46 $\pm$ 0.18 & 0.89 $\pm$ 0.09 & -- & -- & 3 \\
            \hline
\end{tabular}
}
\end{table*}

\subsection{Mass functions of open clusters}
\label{sect:ms}

\begin{figure}
	\includegraphics[width=\columnwidth]{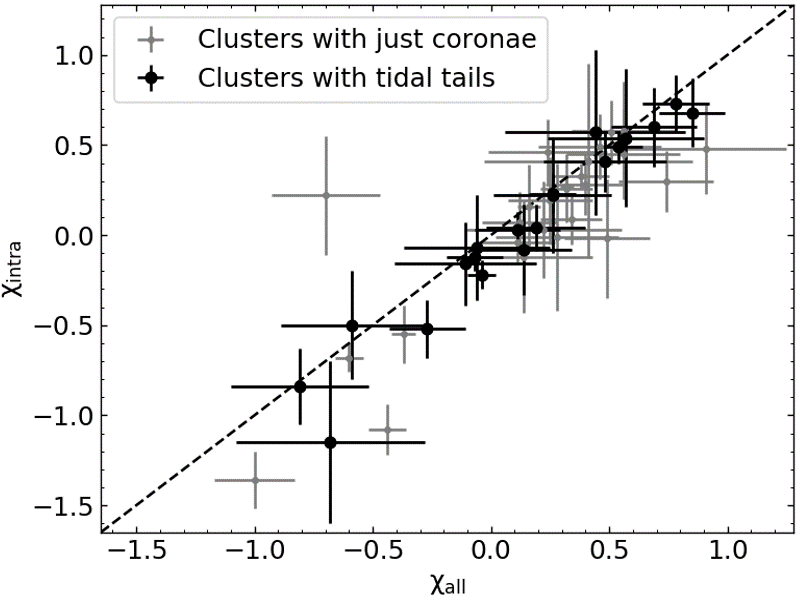}
	\includegraphics[width=\columnwidth]{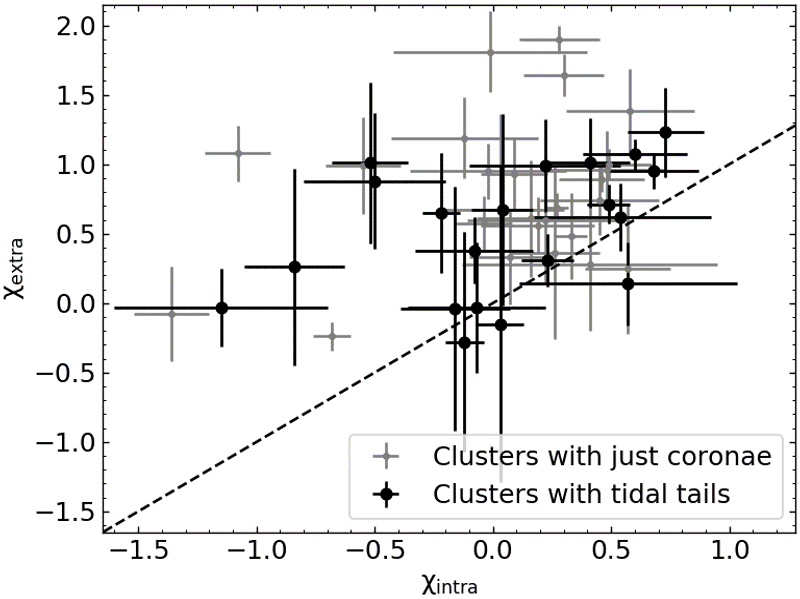}
    \caption{[Top] $\rm\chi_{intra}$ against $\rm\chi_{all}$ for all the clusters (black: having tidal tails; grey: having only a corona). [Bottom] $\rm\chi_{extra}$ against $\rm\chi_{intra}$ for all the clusters. }
    \label{fig:seg}
\end{figure}

In addition to Teutsch~35, we fit mass functions also for the other 45 open clusters {to cluster members more massive than 0.5~M$_{\sun}$} as in Section~\ref{sect:mf}. Given the range of distances and extinctions for the clusters studied in this work, the mass of 0.5~M$_{\sun}$ corresponds to a wide range of G-band magnitudes. We, however, limit ourselves to fitting the mass function to stars brighter than G=18.5~mag (where the cluster membership contamination and completeness are secure; see Appendix~\ref{sect:comp}) for the clusters where 0.5~M$_{\sun}$ corresponds to fainter G-band magnitudes. The fitted mass function slopes for the entire cluster, $\rm\chi_{all}$, for the intra-tidal members, $\rm\chi_{intra}$ and for the extra-tidal members, $\rm\chi_{extra}$ have been noted in Table~\ref{table:seg}. 

Figure~\ref{fig:seg} [top] shows $\rm\chi_{intra}$ against $\rm\chi_{all}$ for all the clusters, separately for the clusters with tidal tails (black) and those only with a corona (grey). The two quantities follow the 1:1 line within error with few exceptions, having no clear difference between the clusters with tidal tails and those with only a corona. This tight correlation also implies that the fitted mass function within the tidal radius is a good approximation for the mass function of the entire cluster. This result confirms that the estimated mass function using cluster members within the inner regions of open clusters (generally close to or within its tidal radius) as carried out in earlier studies \citep[E.g.][]{Bhattacharya17a,Bhattacharya17b} prior to \textit{Gaia} DR2 is still a valid approximation for the entire cluster. For the clusters studied here, both $\rm\chi_{intra}$ and $\rm\chi_{all}$ values are flatter than the Salpeter slope including uncertainties (see Table~\ref{table:seg}).

Figure~\ref{fig:seg} [bottom] shows $\rm\chi_{extra}$ against $\rm\chi_{intra}$ for all the clusters, separately for the clusters with tidal tails (black) and those only with a corona (grey). The two quantities either follow the 1:1 line within error or $\rm\chi_{extra}$ is steeper than $\rm\chi_{intra}$. This implies that open clusters have the same mass function slope in the outskirts as the inner regions (where $\rm\chi_{extra} \approx \chi_{intra}$) or they have a steeper slope in the outskirts (where $\rm\chi_{extra} > \chi_{intra}$). The effect is independent of whether the cluster has identified tidal tails or just a corona. This implies that though inner regions of the cluster may be experiencing mass segregation, with mass function slopes flatter than the Salpeter value, the effect may not be as prominent beyond the tidal radius, where the mass function slopes can be relatively steeper (see Table~\ref{table:seg}). We also calculate the mass function for the tidal tails where possible as noted in Table~\ref{table:seg}. The tidal tails generally have very few stars with the mass function slopes fitted having large uncertainty, though they typically have mass function slopes more positive than $\rm\chi_{intra}$. {We reiterate that since ML-MOC would not be able to identify tail members out to large distances \citep{Boffin22}, the fitted mass function of the tail members is highly uncertain.}

\subsection{Minimum spanning tree and mass segregation classes}
\label{sect:mst}

\begin{figure*}
	\includegraphics[width=\textwidth]{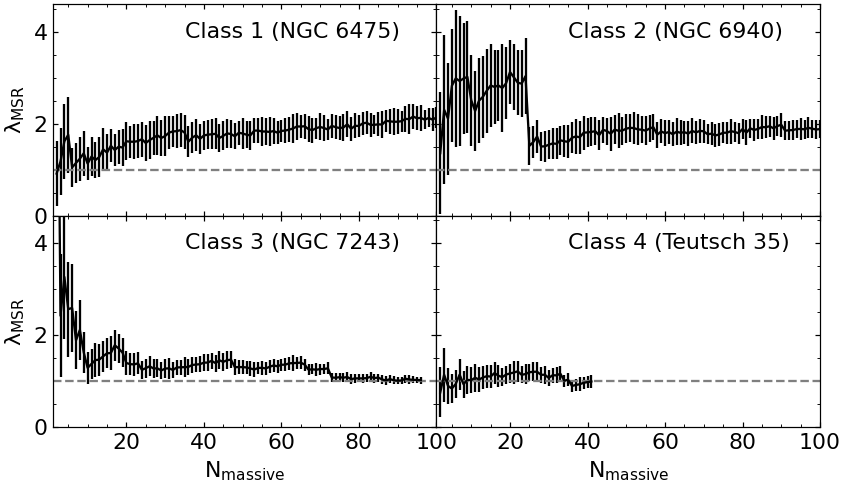}
    \caption{Example $\rm \lambda_{MSR}$ vs. N$\rm_{massive}$ profiles of open clusters of the four assigned mass segregation classes. The error bar shows the standard deviation in $\rm \lambda_{MSR}$ values at each N$\rm_{massive}$. The dashed grey line shows the $\rm \lambda_{MSR}=1$ value which corresponds to no mass segregation. }
    \label{fig:mst}
\end{figure*}

While the deviation of the mass function slope from the Salpeter value is an indicator of mass segregation, the degree of mass segregation can be quantified in open clusters following the prescription by \citet{Allison09} which utilizes minimum spanning trees (MSTs\footnote{{An MST of a set of points is the path connecting all the points, with the shortest possible path length and without any closed loops.}}). The method calculates the mass segregation ratio ($\rm \lambda_{MSR}$) of a cluster by comparing the length of the MST ($\rm l_{massive}$) of the most massive members (numbering N$\rm_{massive}$) with the length of the MST of a set of the same number of randomly chosen member stars ($\rm l_{random}$). $\rm \lambda_{MSR}$ is thus described by the following equation:
\begin{equation}
    $$\rm \lambda_{MSR} (N_{massive}) = \frac{<l_{random}>}{l_{massive}} \pm \frac{\sigma(l_{random})}{l_{massive}}$$
    \label{eq:mst}
\end{equation}
where $\rm<l_{random}>$ and $\rm\sigma(l_{random})$ refer to the mean and standard deviation of $\rm l_{random}$ calculated in N random samples of cluster members. In this work, we sample N$\rm_{massive}$ random cluster members N = 100 times. Only those cluster members brighter than G~=~18.5~mag are sampled in order to avoid completeness issues (see Appendix~\ref{sect:comp} for details). The $\rm \lambda_{MSR}$ value thus calculated varies with the chosen N$\rm_{massive}$ value. If a cluster is mass segregated, its $\rm \lambda_{MSR}$ value would be greater than 1, with the value increasing with higher degrees of mass segregation. If more cluster members feel the effect of mass segregation, then $\rm \lambda_{MSR}$ value would remain greater than 1 for higher N$\rm_{massive}$ values.

We calculate the $\rm \lambda_{MSR}$ values for each of the 46 clusters studied in this work limiting the N$\rm_{massive}$ value to the number of cluster members brighter than G~=~14~mag (this generally covers ~3--4 mag below the brightest most massive member even in the most distant clusters in our sample) up to a maximum N$\rm_{massive}=100$. We find that the $\rm \lambda_{MSR}$ vs. N$\rm_{massive}$ profiles of the open clusters can be visually classified into four general classes (noted in Table~\ref{table:seg}). An example of a cluster for each of these classes is shown in Figure~\ref{fig:mst} and we describe each class as follows:
\begin{itemize}
\item \textit{Class 1}: The $\rm \lambda_{MSR}$ value is low (generally below 1) for small N$\rm_{massive}$ values but increases with increasing N$\rm_{massive}$ values remaining above 1 (including the 1~$\sigma$ uncertainty) till the maximum measured N$\rm_{massive}$ value. NGC~6475 is such a cluster whose $\rm \lambda_{MSR}$ vs. N$\rm_{massive}$ profile is shown in Figure~\ref{fig:mst}. There are 6 clusters in our sample which show such profiles, 3 of which have tidal tails. The mass segregation effect seems to be felt for the largest number of cluster members.
\item \textit{Class 2}: The $\rm \lambda_{MSR}$ value is close to its highest value for small N$\rm_{massive}$ values but then decreases with increasing N$\rm_{massive}$ values. However, the $\rm \lambda_{MSR}$ value eventually plateaus to a value above 1 (including the 1~$\sigma$ uncertainty) till the maximum measured N$\rm_{massive}$ value. NGC~6940 is such a cluster whose $\rm \lambda_{MSR}$ vs. N$\rm_{massive}$ profile is shown in Figure~\ref{fig:mst}. There are 16 clusters in our sample which show such profiles, 5 of which have tidal tails. As for Class 1 clusters, the mass segregation effect seems to be felt for a large number of cluster members.
\item \textit{Class 3}: As for Class 2 clusters, the $\rm \lambda_{MSR}$ value is close to its highest value for small N$\rm_{massive}$ values but then decreases with increasing N$\rm_{massive}$ values. However, the $\rm \lambda_{MSR}$ value eventually plateaus to a value around 1 (including the 1~$\sigma$ uncertainty) till the maximum measured N$\rm_{massive}$ value. NGC~7243 is such a cluster whose $\rm \lambda_{MSR}$ vs. N$\rm_{massive}$ profile is shown in Figure~\ref{fig:mst}. There are 19 clusters in our sample which show such profiles, 7 of which have tidal tails. The mass segregation effect seems to be felt only for the most massive cluster members.
\item \textit{Class 4}: These clusters show $\rm \lambda_{MSR}$ values around 1 (including the 1~$\sigma$ uncertainty) for nearly all measured N$\rm_{massive}$ values. Teutsch~35 is such a cluster whose $\rm \lambda_{MSR}$ vs. N$\rm_{massive}$ profile is shown in Figure~\ref{fig:mst}. There are 5 clusters in our sample which show such profiles, all of which have tidal tails. These clusters do not show mass segregation.
\end{itemize}

It is not clear whether Class 1 or Class 2 host the most mass segregated clusters, but broadly speaking they have the clusters where the mass segregation effect is experienced by most members. Class 3 clusters have mass segregation for just the most massive members while Class 4 clusters are non-mass segregated. It is intriguing that all Class 4 clusters in this work host tidal tails. {One may naively interpret} that there is a $78.27 \pm 21.72$\% probability (95\% binomial confidence interval calculated for finding all 5 clusters with tidal tails) that all non-mass segregated clusters have tidal tails. Since such a scenario is unlikely for all non-mass segregated clusters, we speculate that our initial selection of elongated clusters in 2D space (Section~\ref{sect:data}) is responsible for preferentially selecting only those non-mass segregated clusters with tidal tails. It is possible that an elongated cluster that is not mass segregated would only seem elongated because of its tidal tails while elongated clusters that are mass segregated would have a more populated halo of lower mass stars whose 2D spatial distribution may appear elongated. Thus all 5 Class 4 clusters hosting tidal tails may just stem from our initial selection of elongated clusters, {implying that such a selection of non-mass segregated clusters would yield a $78.27 \pm 21.72$\% probability of such clusters having tidal tails.} 

\subsection{Mass segregation and cluster age}
\label{sect:mst_age}

\begin{figure}
	\includegraphics[width=\columnwidth]{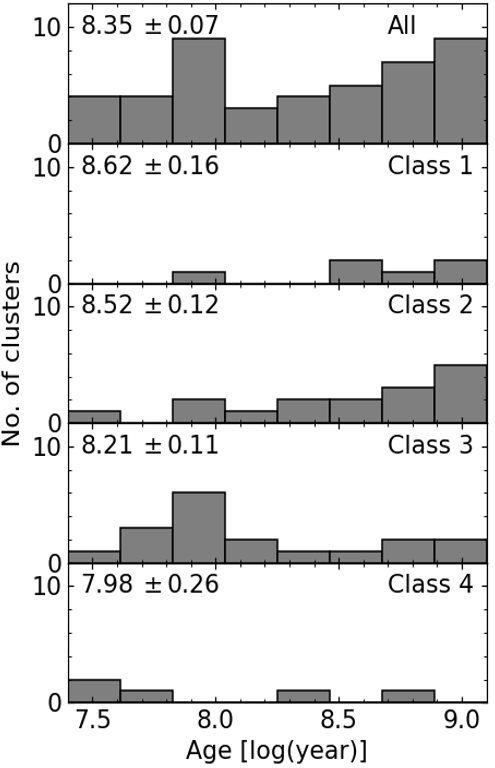}
    \caption{Histogram of Ages of the open clusters studied in this work, for the full sample in the top panel and divided by the assigned mass segregation class in subsequent panels. The {mean age} for each class is also marked.}
    \label{fig:class_age}
\end{figure}

Mass segregation in open clusters is believed to increase with age with old clusters thought to be more frequently mass segregated than young clusters \citep{Kroupa95,Dib18}. Figure~\ref{fig:class_age} shows the histogram of cluster ages for the full sample of open clusters as well as the subsamples for each class. The mean age {($\pm$ error on the mean)} for each class is also marked. As expected from the $\rm \lambda_{MSR}$ vs. N$\rm_{massive}$ profiles, the most-mass segregated Class 1 and Class 2 clusters are the oldest with mean ages of $8.62 \pm 0.16$~{log(year)} and $8.52 \pm 0.12$~{log(year)} respectively. It is unclear whether Class 1 or Class 2 open clusters are older as their mean ages overlap. Class 3 clusters, less mass segregated than Class 2, are younger on average with a mean age of $8.21 \pm 0.11$~{log(year)}. The non-mass segregated Class 4 clusters with a rather uncertain mean age $7.98 \pm 0.26$~{log(year)} of are younger than Class 1 and Class 2 clusters but overlap in mean age with Class 3 cluster. 

{Using an Anderson-Darling test (AD-test; \citealt{ksample_ADtest}; see Section~3.4 in \citealt{Bh21} for a detailed description), we can statistically compare if the cluster ages in the different classes are drawn from the different parent distributions. The test requires at least 8 clusters in each class to make an accurate comparison. Hence, we restrict our comparison to the ages of the most mass segregated clusters (22 clusters in Class 1 and Class 2 combined) with that of the relatively less mass-segregated Class 3. With a significance of 0.03 (less than 0.05 thereby rejecting the null hypothesis that the two samples are drawn from the same parent distribution), the AD-test confirms that the Class 1 and Class 2 clusters have a statistically distinct age (older on average) than Class 3 clusters. We thus statistically find that more mass segregated clusters are older on average than less mass segregated ones.} 

The variation of the degree of mass segregation with age was also studied by \citet{Tarricq22} for their open clusters (see their Section~5) using the same $\rm \lambda_{MSR}$ prescription. They had defined  $\rm \lambda_{10}$ as the MSR for the ten most massive members to compare the degree of mass segregation between clusters. They had found no correlation of  $\rm \lambda_{10}$ with age. Given the variation of the $\rm \lambda_{MSR}$ values with N$\rm_{massive}$, we do not think it is appropriate to compare the degree of mass segregation between clusters by restricting the comparison to the $\rm \lambda_{MSR}$ for any one N$\rm_{massive}$ value. Finding an appropriate quantifiable comparison between these profiles for different clusters is beyond the scope of this work. The broad division into classes is sufficient to show {hints} of mass segregation increasing with age (Figure~\ref{fig:class_age}).

\subsection{Mass segregation and mass function slope}
\label{sect:mst_mf}

\begin{figure}
	\includegraphics[width=\columnwidth]{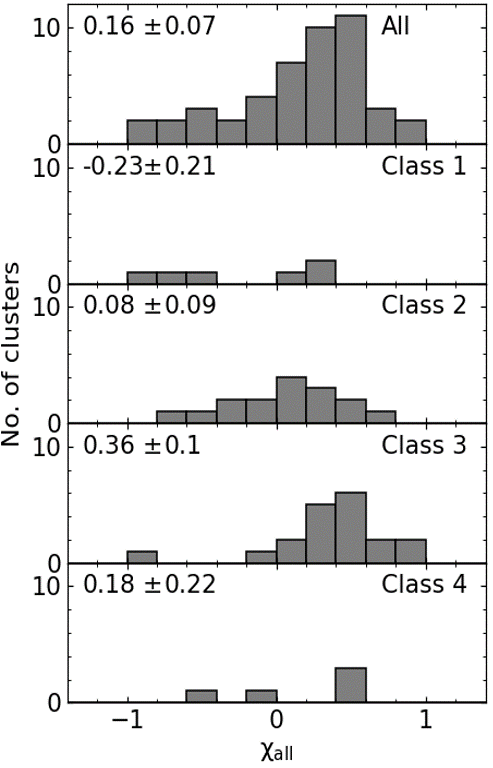}
    \caption{{Histogram of $\rm \chi_{all}$ values of the open clusters studied in this work, for the full sample in the top panel and divided by the assigned mass segregat$\rm \lambda_{MSR}$ vs. N$\rm_{massive}$ profiles of open clusters of the four assigned mass segregation classes.ion class in subsequent panels. The mean $\pm$ standard deviation of $\rm \chi_{all}$ values for each class is also marked.}}
    \label{fig:class}
\end{figure}

The deviation of the mass function slope from the Salpeter value has long been held as an indicator of mass segregation (Section~\ref{sect:mf}). We compare the distribution of the $\rm \chi_{all}$ values of the open clusters studied in this work for the full sample as well as the sub-samples of four mass segregation classes in Figure~\ref{fig:class}. The mean of $\rm \chi_{all}$ values for each class is also marked. As expected from the $\rm \lambda_{MSR}$ vs. N$\rm_{massive}$ profiles, {the more mass segregated Class 1 clusters also have the flattest mass function slopes, most deviant from the Salpeter values. The similarly more mass segregated Class 2 clusters have a mean $\rm \chi_{all}$ value overlapping within error with the Class 1 clusters. Class 3 clusters, less mass segregated than Class 2, have a mean $\rm \chi_{all}$ value relatively closer to the Salpeter value. The non-mass segregated Class 4 clusters have a uncertain mean $\rm \chi_{all}$ value that overlaps within error with the other three classes.  }

{We again utilize the AD-test to compare if the distribution of mass function slopes in the more mass segregated Class 1 and Class 2 clusters is statistically distinct from that in the less mass segregated Class 3 clusters. With a significance of 0.003, the AD-test confirms that the Class 1 and Class 2 clusters have a statistically distinct mass function slope (lower on average) than Class 3 clusters. We thus statistically find that more mass segregated clusters have a larger deviation of the mass function slope from the Salpeter value on average than less mass segregated ones. However, the amount of deviation of the mass function slope of an individual cluster from the Salpeter value (given the broad distribution of $\rm\chi_{all}$ values in each class in Figure~\ref{fig:class}) is not a measure of degree of mass segregation.} Furthermore, given the $\rm \chi_{all}$ values of Class 4 clusters, the mass function slope's deviation from the Salpeter value is also not sufficient to claim mass segregation.

\subsection{Initial cluster mass}
\label{sect:inimass}

\begin{figure}
	\includegraphics[width=\columnwidth]{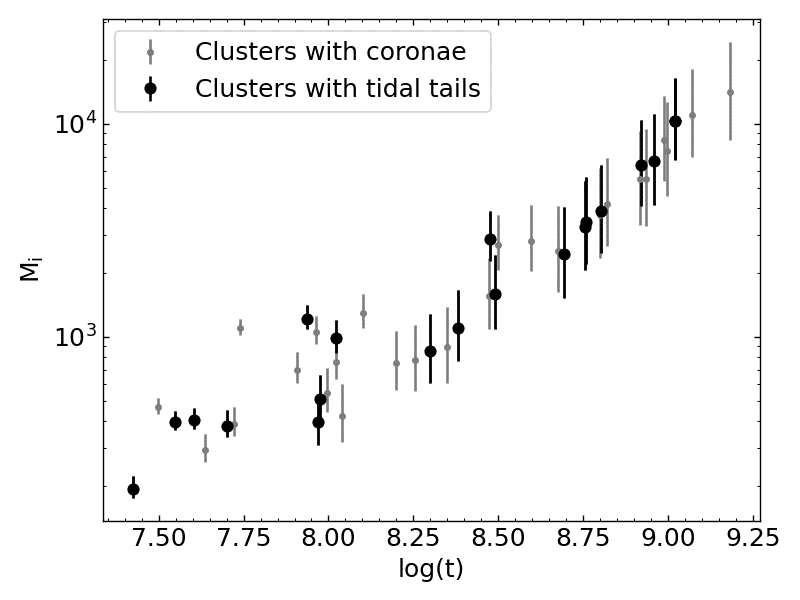}
	\includegraphics[width=\columnwidth]{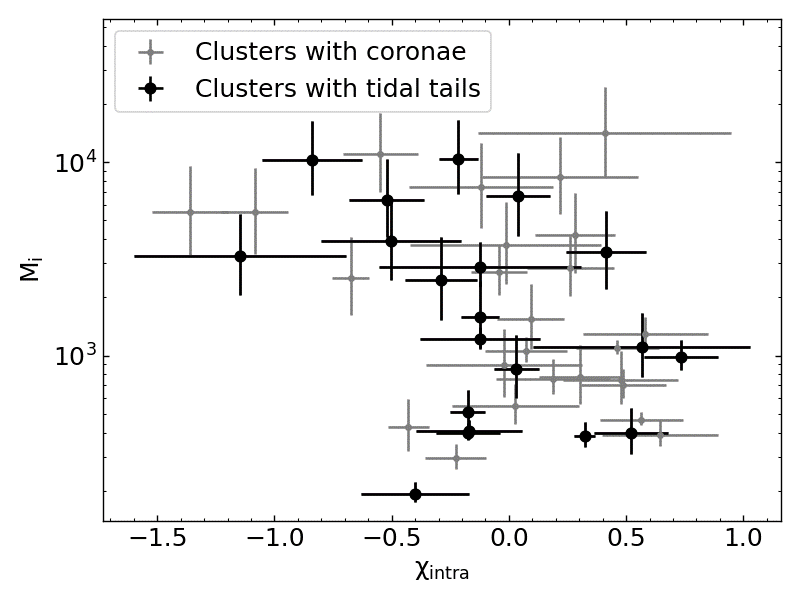}
    \caption{[Top] $\rm M_{i}$ against the logarithm of cluster age for all the clusters (black: having tidal tails; grey: having only a corona). [Bottom] $\rm M_{i}$ against $\rm\chi_{intra}$ for all the clusters. }
    \label{fig:mi}
\end{figure}

The mass functions of all {46} open clusters presented here, 20 of which have extended tidal features are a telltale sign of their dissolution. Environmental effects like tidal shocks due to close interactions with giant molecular clouds, spiral arms, the Galactic disc and, in general, interactions with the Galactic tidal field or internal dynamical effects like two-body relaxation lead to mass-loss in an open cluster \citep{Dalessandro15}. We can estimate the initial mass ($\rm M_{i}$) of an open cluster which has lost mass due to interaction with the Galactic tidal field following the analytical prescription by \citet{Lamers05}, as was done for NGC~752 by \citet{Bhattacharya21}. $\rm M_{i}$ is estimated with the following relation:
\begin{equation}
    $$\rm M_{i} = \rm \bigg[\bigg(\frac{M_{C}}{M_{\sun}}\bigg) ^{\gamma} + \frac{\gamma t}{t_{0}} \bigg]^{\frac{1}{\gamma}} [1-q_{ev}(t)]^{-1}$$
    \label{eq:mini}
\end{equation}
where t$\rm_{0}$ is the dissolution time-scale parameter and {M$\rm_{C}$ is the present cluster mass}. By comparing the distribution of mass and age of OCs in the solar neighbourhood with theoretical predictions, \citet{Lamers05} obtained t$\rm_{0}=3.3_{-1.0}^{1.4}$ Myr. $\rm \gamma$ is a dimensionless index which depends on the cluster initial density distribution. We adopt $\rm \gamma=0.62$ which is the typical value for open clusters \citep{Lamers05, Dalessandro15}. The function q$\rm_{ev}$(t) describes the mass-loss due to stellar evolution and can be approximated by the following analytical formula:
\begin{equation}
    $$\rm log_{10}(q_{ev}(t)) = \rm (log_{10}(t)-a)^{b}+c$$
    \label{eq:qev}
\end{equation}
where a~=~$7.0$, b~=~$0.255$ and c~=~$-1.805$, appropriate for all our clusters that have near-solar metallicity, except Alessi~6 for which a~=~$7.03$, b~=~$0.26$ and c~=~$-1.8$ \citep[see Table 1 in][]{Lamers05}. Given that the unobserved mass would be small for all our clusters with only the least massive stars remaining unobserved we assume M$\rm_{C}=$~m$\rm_{p}$. Furthermore, the uncertainty in $\rm M_{i}$ is dominated by the uncertainty in t$\rm_{0}$ with the uncertainty in M$\rm_{C}$ having a negligible effect, as was seen for NGC~752 \citep{Bhattacharya21}. We thus ignore the uncertainty in M$\rm_{C}$ in calculating $\rm M_{i}$. Plugging the values in to Equations~\ref{eq:mini}~\&~\ref{eq:qev}, we find $\rm M_{i}$ for all our clusters, noted in Table~\ref{table:clus}. Gulliver~44 has the highest $\rm M_{i}=14108^{10180}_{-5698}$~M$_{\sun}$ having lost $98.77^{0.52}_{-0.84}$\% of its mass, while NGC~3228 has the lowest $\rm M_{i}=193^{28}_{-18}$~M$_{\sun}$ having lost just $37.44^{8.55}_{-6.44}$\% of its mass. Notably, Gulliver~44, IC~4756, NGC~1662, NGC~2423, NGC~2527, NGC~5822 and NGC~6940 are candidates for being the remnants of Young Massive Clusters (which have stellar masses $\gtrsim 10^{4} \rm M_{\sun}$) such as those observed in the Milky Way and other galaxies \citep[see review by][]{Zwart10}. 

Figure~\ref{fig:mi} [top] shows the variation of $\rm M_{i}$ with the logarithm of cluster age. Older clusters are also clearly seen to have higher initial masses, as expected from Equation~\ref{eq:mini}. There is no clear difference between clusters with tidal tails and those with only a corona. Figure~\ref{fig:mi} [bottom] shows the variation of $\rm M_{i}$ with $\rm\chi_{intra}$. The clusters having $\rm M_{i}<2000$~M$_{\sun}$ have higher $\rm\chi_{intra}$ and thus are relatively less mass segregated. Clusters with $\rm M_{i}>2000$~M$_{\sun}$ tend to cover a large range in $\rm\chi_{intra}$, indicating that while all the clusters are mass segregated there is no clear relation between $\rm\chi_{intra}$ and the amount of mass segregation.

\subsection{Notes on individual clusters}
\label{sect:notes}
\begin{itemize}
    \item \textit{The Pleiades}: \citet{Li21} identified extended features beyond the tidal radius of the Pleiades but classified these features as early-stage tidal features, with much shorter tails than those found for the similarly aged Blanco~1 which extend out to $\sim$50--60~pc \citep{Zhang20,Pang21}. We identify the leading and lagging tail of the Pleiades to be 41.19~pc and 34.5~pc long respectively, within range of that of Blanco~1. {Figure~\ref{fig:orbit_append} for the Pleiades shows that the leading and lagging tidal tails respectively point away from and towards the direction intermediate between the cluster orbit and the GC direction in all three panels. However, conspicuously, there are extra-tidal members other than the tidal tails which point towards the positive Z direction from the cluster center visible in the X-Z and Y-Z panels for the Pleiades in Figure~\ref{fig:orbit_append}. This stream of stars does not seem to be tidally induced as per the criteria discussed in Section~\ref{sect:tt}. It corresponds to the direction of the extended tail of the cluster identified by \citet{Lodieu19} though we do not find any cluster members in the positive Z direction from the cluster center as was found by \citet{Lodieu19}. }
    \item {\textit{Alpha Persei}: \citet{Nikiforova20} identified tidal tails in Alpha Persei from stellar overdensities, but were not able to determine members for the tails. In this work, we are also unable to identify the tail members and detect just the cluster corona, same as \citet{Lodieu19}. }
    \item {\textit{Candidate clusters with tidal tails from \citet{Tarricq22}}: As previously mentioned, \citet{Tarricq22} identified extended features in 72 open clusters terming them tidal tails though they  did not investigate whether the extended features were tidally formed. 10 of their candidate clusters with tidal tails are present in our sample. For 5 of them (namely NGC~2527, NGC~5822, NGC~6940, M~39, and Teutsch~35), we confirm the tidal nature of their tails from the distribution of their identified members with respect to the cluster orbit in the galactocentric phase space (Figure~\ref{fig:orbit_append}). For the other 5 (namely Alessi~Teutsch~3, NGC~1039, NGC~1528, NGC~2423, and NGC~2546), we find that the extra-tidal members do not show any extended features which point to or away from the direction intermediate between the cluster orbit and direction of the GC (Figure~\ref{fig:orbit_notails}). Thus these clusters do not have tidal tails identified in this work.}
    \item \textit{IC~4756}: \citet{Ye21} claim tidal tails in IC~4756 as it appears extended in the X-Y plane of the galactocentric phase space, similar to the tails of the Hyades \citep{Meingast19} and Praesepe \citep{Roser19b}. However the extension in the X-Y plane {may simply be} due to the line-of-sight extension at the distance of IC~4756 and not due to tidal tails\footnote{{IC 4756, however, may have tidal tails based on the dissipation of their spatially identified tail members using measured Gaia DR3 astrometry (Xianhao Ye, private communication).}}. In this work, we identify the extended corona of this cluster for the first time. 
    \item \textit{Previously identified clusters with coronae}: Among the 20 clusters having tidal tails identified in this work, IC~2602, M~39 and Pleiades have already been identified by \citet{Meingast21} to have a corona although they do not explore the tidal nature in these clusters. \citet{Pang21} searched for tidal tails in IC~2602 but did not identify them. In NGC~2451B, they identified no tidal tails but did identify an extended feature that may be a possible remnant of a star-forming filament ({or formed from early stage gas expulsion in the cluster as predicted by \citet{Tutukov78}}). We also identify this extended structure, most clearly visible as the extra-tidal members of NGC~2451B not belonging to either tail in the galactocentric phase space in Figure~\ref{fig:orbit_append}. Thus in this work, we identify the tidal tails of IC~2602 and NGC~2451B for the first time even though their extended corona was known. IC~4665 had already been observed by \citet{Pang21} to have a corona, which is also what we find.
    \item \textit{Newly discovered extended coronae and tidal tails}: {The tidal tails of all other clusters mentioned have been discovered in this work totalling 14 clusters, with 5 more identified by \citet{Tarricq22} being confirmed to have tidal tails in this work. We report just the extended coronae of 26 open clusters.} Notably, NGC~6940 is the most distant cluster with tidal tails discovered from \textit{Gaia} data at over 1~kpc away. Alessi~6 in particular has an elongated feature in addition to its tidal tails (see its extra-tidal members in Figure~\ref{fig:orbit_append}). Given the cluster has an age of log(t)~=~8.803, it is unlikely that this elongated feature is associated with the remnants of a filament. However, it is possible that this feature belongs to a stream associated with Alessi~6, as was found for Coma Berenices \citep{Tang19} and Alpha Persei \citep{Nikiforova20}. 
\end{itemize}

\section{Summary and conclusion}
\label{sect:con}
Selecting a sample of open clusters having elongated morphology \citep[determined by][]{Hu21}, we utilise the robust membership determination algorithm, ML-MOC, to identify their member stars in the deep \textit{Gaia} EDR3 data. {46} open clusters were found to have an extended corona of stars beyond their tidal radius. {We computed the mass function slopes (Section~\ref{sect:mf}) for all the clusters separately for all members, the intra-tidal members, the extra-tidal members and the tails where possible. The mass function slopes for all members and just the intra-tidal members were flatter than the Salpeter value for all clusters.} The mass function slope for the intra-tidal region was shown to be a good approximation of the mass function slope of the entire cluster. We also estimated the initial mass for the open clusters, which was found to increase linearly with age. Furthermore, clusters with low $\rm M_{i}$ had a relatively higher $\rm\chi_{intra}$ while those with high $\rm M_{i}$ spanned a wide range in $\rm\chi_{intra}$. 

We further computed the $\rm \lambda_{MSR}$ vs. N$\rm_{massive}$ profiles (Section~\ref{sect:mst}) of the open clusters using the MST-based prescription described in \citet{Allison09}. From these profiles, we can divide our sample into four assigned mass segregation classes, {with Class 1 and Class 2 clusters being more mass segregated, being the oldest and showing the largest deviation of the mass function slope from the Salpeter slope on average. Class 3 clusters are relatively less mass segregated, having lower age on average and lower average mass function slope deviation from the Salpeter value compared to Class 1 and Class 2 clusters. Class 4 clusters are non-mass segregated with a younger mean age but consistent within errors with Class 3 clusters.} 41 of the 46 open clusters studied in this work exhibit mass segregation to some degree. We note that the mass function slope's deviation from the Salpeter value is not a measure of an individual cluster's degree of mass segregation and is also not a sufficient condition for mass segregation in a cluster (Section~\ref{sect:mst_mf}).

20 of the open clusters {have} tidal tails based on their relative position to the cluster orbit direction, discussed in Section~\ref{sect:orbit}. 14 of them have their tidal tails identified for the first time in this work with 5 more candidates {whose tail alignment with the cluster orbit are confirmed} in this work. We find more extended tails for the Pleiades than previous works. We also find that NGC~6940, the most distant open cluster with tidal tails found till date, has the longest physical distance spanned by its tidal tails in this work. We thus add 19 open clusters to the 13 that were already known to have tidal tails. With future data releases of \textit{Gaia}, we expect to find even more open clusters with tidal tails while also better estimating the spatial distribution of those already known {as well as checking the dissipation of tail members}. 

\section*{Acknowledgements}
We thank the anonymous referee for their comments which helped to improve the manuscript. S. Bhattacharya is funded by the INSPIRE Faculty award (DST/INSPIRE/04/2020/002224), Department of Science and Technology (DST), Government of India. This work presents results from the European Space Agency (ESA) space mission \textit{Gaia}. \textit{Gaia} data are being processed by the Gaia Data Processing and Analysis Consortium (DPAC). Funding for the DPAC is provided by national institutions, in particular the institutions participating in the Gaia MultiLateral Agreement (MLA). This research made use of Astropy-- a community-developed core Python package for Astronomy \citep{Astropy13}, SciPy \citep{scipy}, NumPy \citep{numpy} and Matplotlib \citep{matplotlib}. This research also made use of NASA’s Astrophysics Data System (ADS\footnote{\url{https://ui.adsabs.harvard.edu}}).

\section*{Data Availability}
The data underlying this article are publicly available at \url{https://archives.esac.esa.int/gaia}. {The identified members of the open clusters are noted in Table~\ref{table:memb} and will be available in full at the CDS (\url{https://cds.u-strasbg.fr/})}.



\bibliographystyle{mnras}
\bibliography{ref_oc} 




\appendix


\section{Completeness of \textit{Gaia} EDR3 data and contamination of ML-MOC membership determination}
\label{sect:comp}

\begin{figure}
	\includegraphics[width=\columnwidth]{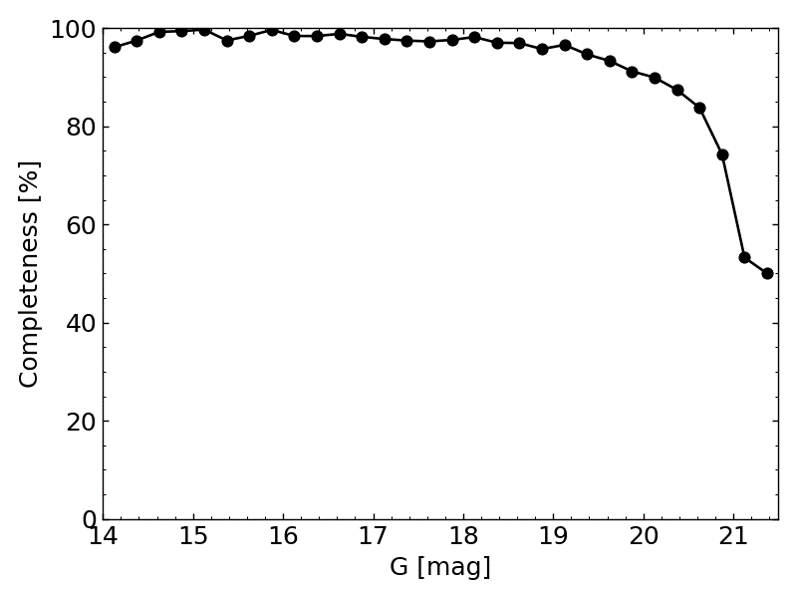}
    \caption{Percentage completeness of the \textit{Gaia} EDR3 catalogue at the position of NGC~2448.}
    \label{fig:comp}
\end{figure}

\begin{figure}
	\includegraphics[width=\columnwidth]{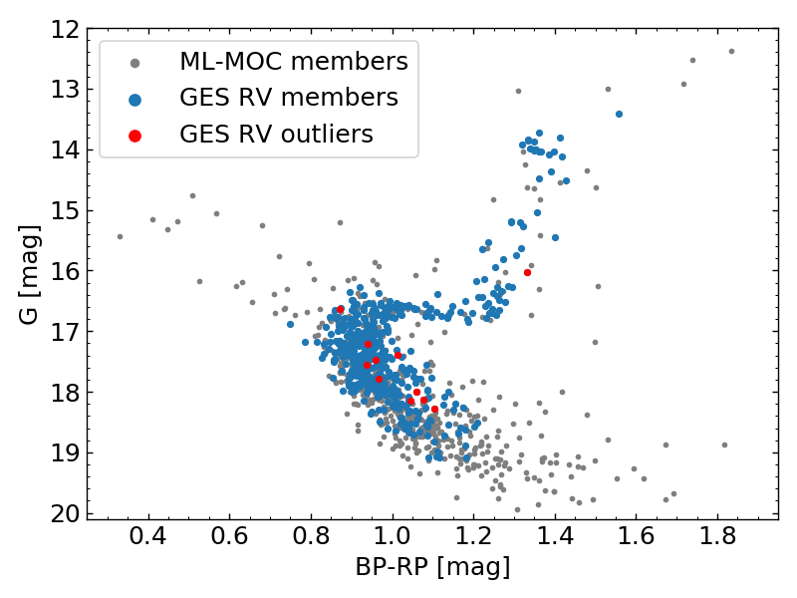}
    \caption{CMD showing the ML-MOC member candidates identified for Berkeley~39 from \textit{Gaia} EDR3 data. Those candidates classified as members using their RV measurements from GES are marked in blue while those classified as outliers are marked in red.}
    \label{fig:cont}
\end{figure}

\textit{Gaia} EDR3 completeness has been explored for globular clusters by comparing the observed \textit{Gaia} EDR3 sources with deeper Hubble Space Telescope data \citep{Fabricius21}. The 90\% completeness limit ranged from G$\sim$16--20 mag depending mostly on stellar density (and the scanning law), being deepest at the less dense outskirts of some of the globular clusters\footnote{see \href{https://gea.esac.esa.int/archive/documentation/GEDR3/Catalogue_consolidation/chap_cu9val/sec_cu9val_947/ssec_cu9val_947_astrometry.html}{\textit{Gaia} EDR3 Documentation Chapter 8.7.1} for examples.}. The number density towards most open clusters is $\sim10^4$--$10^5$ stars per sq. deg. at their densest centers \citep[E.g.][]{Rao21}, even less dense than the outskirts of globular clusters. We explore the completeness of \textit{Gaia} EDR3 data at the position of NGC~2448, as a representative of open clusters in this work. We compare the \textit{Gaia} EDR3 data with the deeper Pan-STARRS-1 DR2 catalogue \citep{Flewelling20} in the same spatial region around this cluster (within 15.07~pc, which is the cluster radius). The Pan-STARRS-1 DR2 (PS1) astrometry is already matched to the \textit{Gaia} reference frame, so we identify PS1 counterparts in the \textit{Gaia} EDR3 data around this cluster. 90\% of PS1 sources are identified down to G$\sim20$~mag with the percentage completeness rapidly declining beyond G$\sim20.5$~mag (Figure~\ref{fig:comp}). While the \textit{Gaia} EDR3 completeness is a function also of the \textit{Gaia} scanning law \citep{Fabricius21}, the position of most of our open clusters near the galactic plane (Figure~\ref{fig:gal}) would mean they occupy the deepest parts of the sky scanned by \textit{Gaia}. As such, the 90\% completeness limit would likely not be much brighter than G$\sim20$~mag for these clusters.

Contamination of ML-MOC selected cluster members in \textit{Gaia} DR2 data has been explored by \citet{Agarwal21} with contamination fractions varying between 5--12\%, with contaminants being mostly present at the faint end. With the improved \textit{Gaia} EDR3 photometry and astrometry, the percentage contamination of ML-MOC members is expected to reduce. A way to check the contamination fraction is to compare the ML-MOC members identified in an open cluster from \textit{Gaia} EDR3 data with RV members determined from deep spectroscopic data. Since none of the open clusters studied in this work have such deep spectroscopic data available publicly, we utilise the ML-MOC members of Berkeley~39 (whose blue straggler population was studied by \citealt{Vaidya20} and \citealt{Rao21} from \textit{Gaia} DR2) which has deep spectroscopic observations from the Gaia ESO survey \citep[GES][]{Bragaglia22} down to G$\sim19.5$~mag. ML-MOC identified 963 members in Berkeley~39 out to $\sim44'$ from the cluster center, 481 of which have RV measurements from GES (limited to within $14'$ from the cluster center). The mean RV of these GES stars is 59.17~km~s$^{-1}$ with $\sigma=5.62$~km~s$^{-1}$. We assume that those sources within 3~$\sigma$ of the mean RV are cluster members while the rest are outliers (its possible for members to have RV beyond this limit especially if they are binary stars but we take this as a conservative estimate). Only 11 stars were identified as outliers this way giving a contamination of 2.3\%. Figure~\ref{fig:cont} shows all the ML-MOC members of Berkeley~39 in grey, those identified as GES RV members in blue and those as GES RV outliers in red. Note the outliers cover a wide magnitude range and are not restricted to the faint end. As such GES RV members are reliably identified also down to G$\sim19.5$~mag. The contamination percentage of ML-MOC members for the clusters studied in this work should also be close to 2.3\% down to G$\sim19.5$~mag, where the Gaia data is also more than 90\% complete.

\section{Identified cluster members and colour magnitude diagrams}
\label{sect:members}

\begin{table*}
\caption{Members identified in each of the 46 open clusters studied in this work with their Gaia EDR3 IDs and some of their parameters. A portion of this table is shown here for guidance; the full table will be made available through the CDS.}
\label{table:memb}
    \centering 
    \adjustbox{max width=\textwidth}{
    \begin{tabular}{ccccccccccc}
    \hline
Cluster name & Gaia EDR3 ID & RA & DEC & $\rm \omega$ & $\rm \mu_{\alpha}cos(\delta)$ & $\rm \mu_{\delta}$ & G & BP & RP & RV\\
 &  & deg & deg & mas & mas/yr & mas/yr & mag & mag & mag & km/s\\
\hline
Alessi 1 & 402506369136008832 & 13.314343 & 49.531599 & 1.415 & 6.363 & -6.391 & 9.83 & 10.33 & 9.15 & -4.19 \\
Alessi 1 & 402506162977579648 & 13.336164 & 49.515789 & 1.43 & 6.636 & -6.382 & 13.2 & 13.5 & 12.75 & -- \\
Alessi 1 & 402508190203269376 & 13.36449 & 49.572677 & 1.411 & 6.36 & -6.435 & 10.51 & 10.63 & 10.28 & -- \\
Alessi 1 & 402505819380310016 & 13.41344 & 49.530571 & 1.403 & 6.535 & -6.368 & 9.85 & 10.23 & 9.29 & -1.83 \\
Alessi 1 & 414516437726588800 & 13.283764 & 49.56353 & 1.443 & 6.444 & -6.507 & 11.44 & 11.6 & 11.16 & -- \\
Alessi 1 & 402505991180022528 & 13.334919 & 49.480474 & 1.474 & 6.483 & -6.368 & 9.82 & 10.32 & 9.15 & -5.23 \\
Alessi 1 & 402500317524014976 & 13.311316 & 49.477524 & 1.445 & 6.514 & -6.656 & 12.75 & 13.0 & 12.34 & -- \\
Alessi 1 & 402507331208797056 & 13.440694 & 49.555027 & 1.411 & 6.537 & -6.463 & 14.72 & 15.15 & 14.12 & -- \\
Alessi 1 & 414509978093147136 & 13.246368 & 49.506712 & 1.435 & 6.548 & -6.421 & 10.81 & 10.96 & 10.53 & -- \\
Alessi 1 & 402507911026190592 & 13.410617 & 49.614802 & 1.369 & 6.477 & -6.351 & 11.69 & 11.87 & 11.38 & -- \\
Alessi 1 & 402507571726963968 & 13.479087 & 49.577303 & 1.426 & 6.482 & -6.531 & 11.24 & 11.38 & 10.97 & -- \\
Alessi 1 & 402502452125956864 & 13.472099 & 49.476324 & 1.363 & 6.406 & -6.382 & 15.63 & 16.19 & 14.93 & -- \\
Alessi 1 & 414510459132125824 & 13.11904 & 49.491571 & 1.382 & 6.446 & -6.419 & 15.25 & 15.74 & 14.6 & -- \\
Alessi 1 & 414508878584171904 & 13.132744 & 49.451849 & 1.424 & 6.484 & -6.401 & 12.61 & 12.86 & 12.21 & -6.44 \\
            \hline
\end{tabular}
}
\end{table*}

\begin{figure*}
	\includegraphics[width=\textwidth]{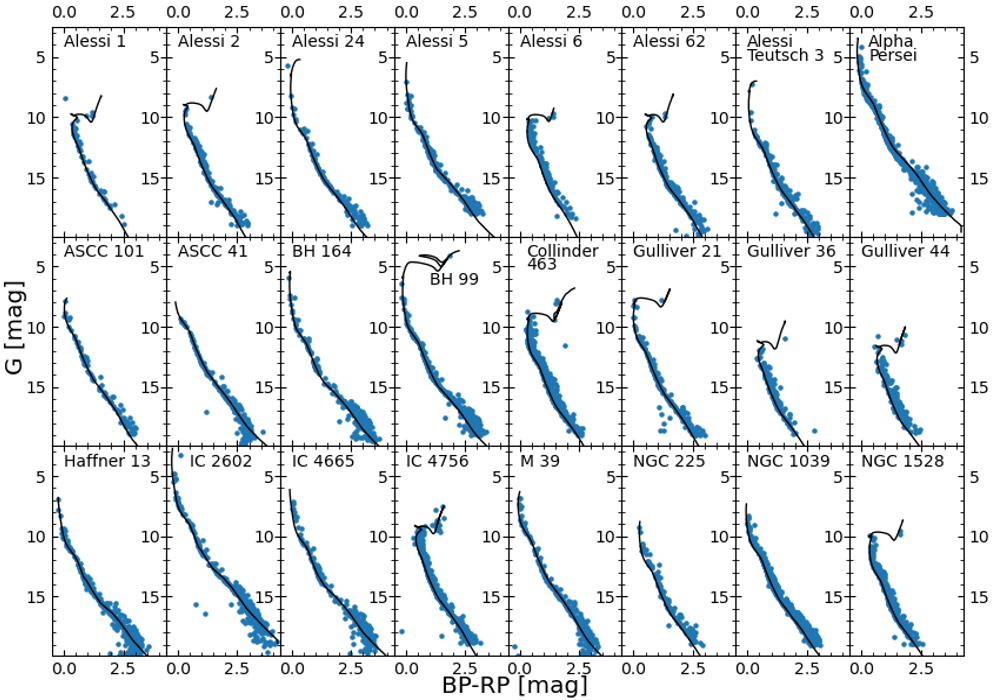}
    \caption{{CMDs of identified members of each open cluster studied in this work. The PARSEC isochrones are plotted in black for the cluster parameters noted in Table~\ref{table:clus}.}}
    \label{fig:cmd_all}
\end{figure*}

\begin{figure*}\ContinuedFloat
	\includegraphics[width=\textwidth]{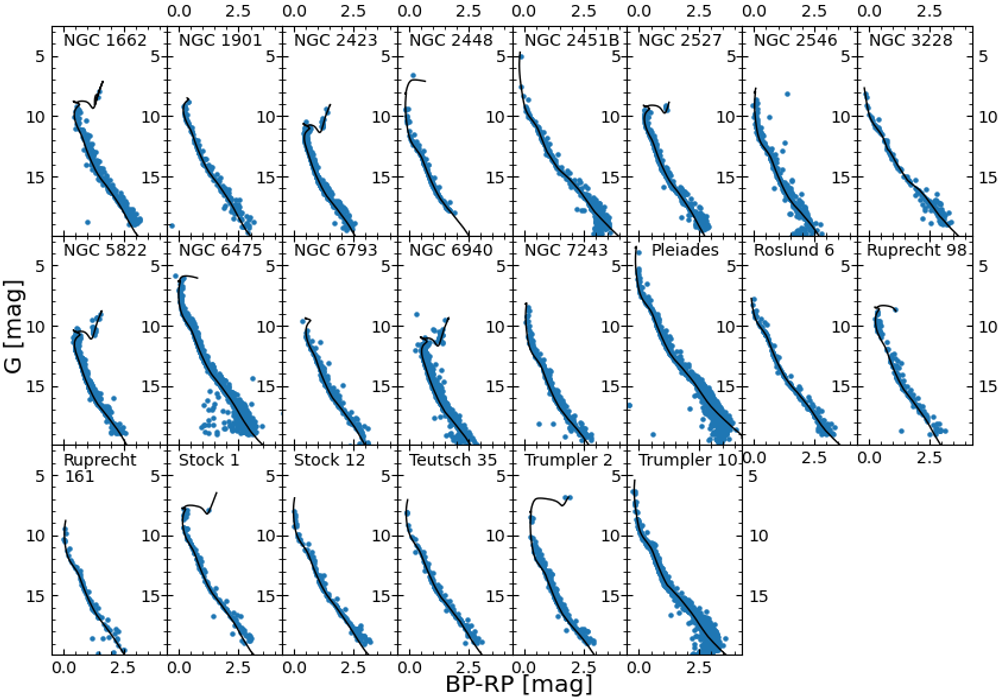}
    \caption{{CMDs of identified members of each open cluster studied in this work. The PARSEC isochrones are plotted in black for the cluster parameters noted in Table~\ref{table:clus}.}}
\end{figure*}

{Members identified by ML-MOC in each of the 46 open clusters are listed in Table~\ref{table:memb} with their Gaia EDR3 IDs, positions, parallax, proper motions, radial velocity and photometric magnitudes. Other measurements for each of these sources have been provided at the Gaia EDR3 archive\footnote{\url{https://archives.esac.esa.int/gaia}}. The CMDs for each of the 46 clusters are shown in Figure~\ref{fig:cmd_all} with PARSEC isochrones plotted for the cluster age, extinction, distance and metallicity, which are noted in Table~\ref{table:clus}.}

\section{Tidal radius overestimation from radial density}
\label{sect:tarricq}

\begin{figure}
	\includegraphics[width=\columnwidth]{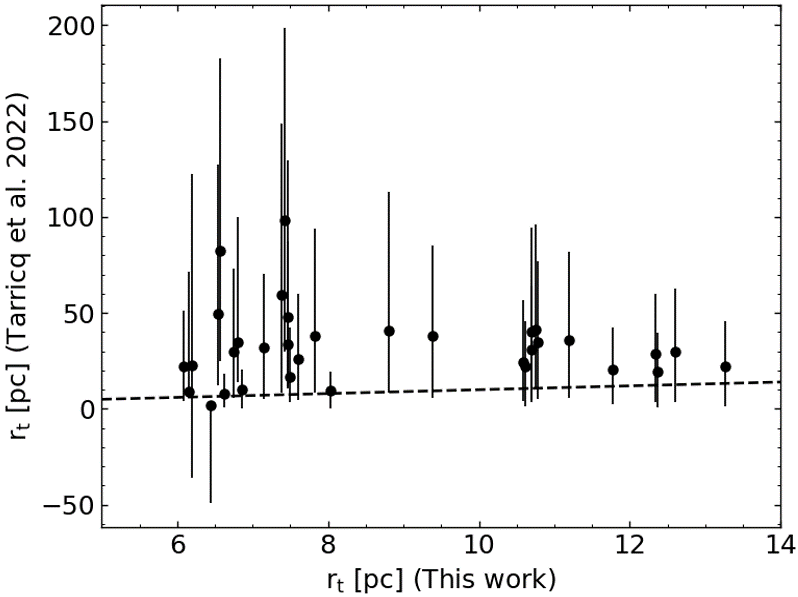}
    \caption{{The r$\rm_{t}$ value obtained by \citet{Tarricq22} in compared to that obtained in this work for the 33 open clusters in common. The 1:1 line is shown as a black dashed line.}}
    \label{fig:tarricq}
\end{figure}

{Fitting the \citet{King66} profile to the radial number density of cluster members (including the members in the corona and extended tails) results in a higher estimate of r$\rm_{t}$ than that measured from the cluster mass radial distribution (see Section~\ref{sect:tr}). The uncertainty on r$\rm_{t}$ is also quite large due to the sparsely populated extra-tidal members at large radii. \citet{Tarricq22} have estimated r$\rm_{t}$ for their studied open clusters from the King profile fit to their identified members. In Figure~\ref{fig:tarricq}, we compare the estimated r$\rm_{t}$ for the 33 open clusters in common between \citet{Tarricq22} and this work. The r$\rm_{t}$ value from \citet{Tarricq22} is significantly larger in most cases compared to that estimated from this work. Such a large r$\rm_{t}$ value would erroneously categorise many tidal tail members as intra-tidal members in some open clusters.}

\section{Tri-dimensional projections in galactocentric coordinates}

\begin{figure*}
	\includegraphics[width=0.83\textwidth]{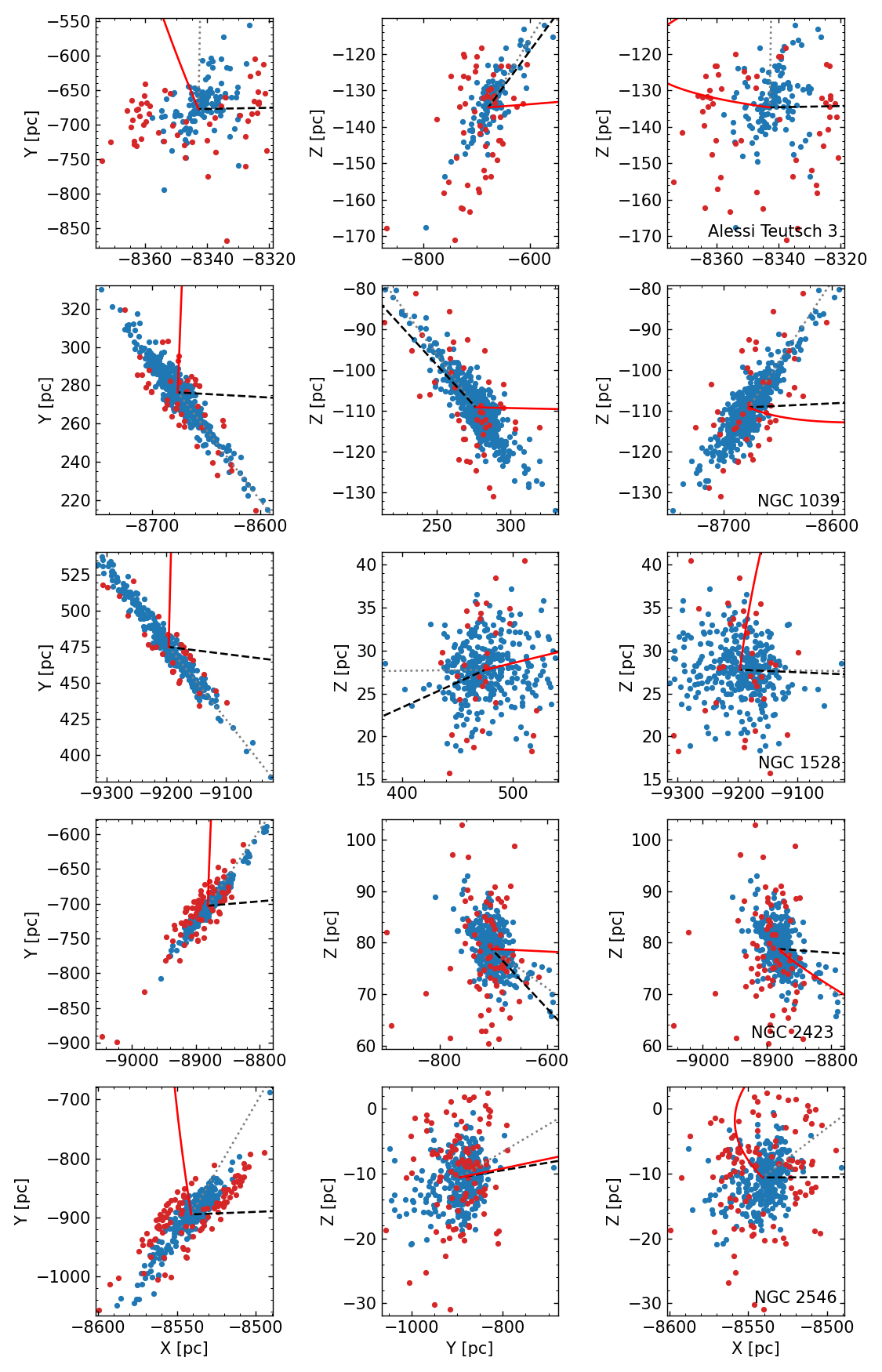}
    \caption{Same as Figure~\ref{fig:orbit} but for the five open clusters identified as having tidal tails in \citet{Tarricq22} but without tidal tails in this work.}
    \label{fig:orbit_notails}
\end{figure*}

{Figure~\ref{fig:orbit_append} shows the Tri-dimensional projections in galactocentric coordinates of 19 open clusters with tidal tails along with their fitted orbits, direction to the GC and the line-of-sight direction. Figure~\ref{fig:orbit_notails} shows the same for the 5 open clusters which were classified as having tidal tails by \citet{Tarricq22} but not found to have tidal tails in this work.} 

\begin{figure*}
	\includegraphics[width=0.83\textwidth]{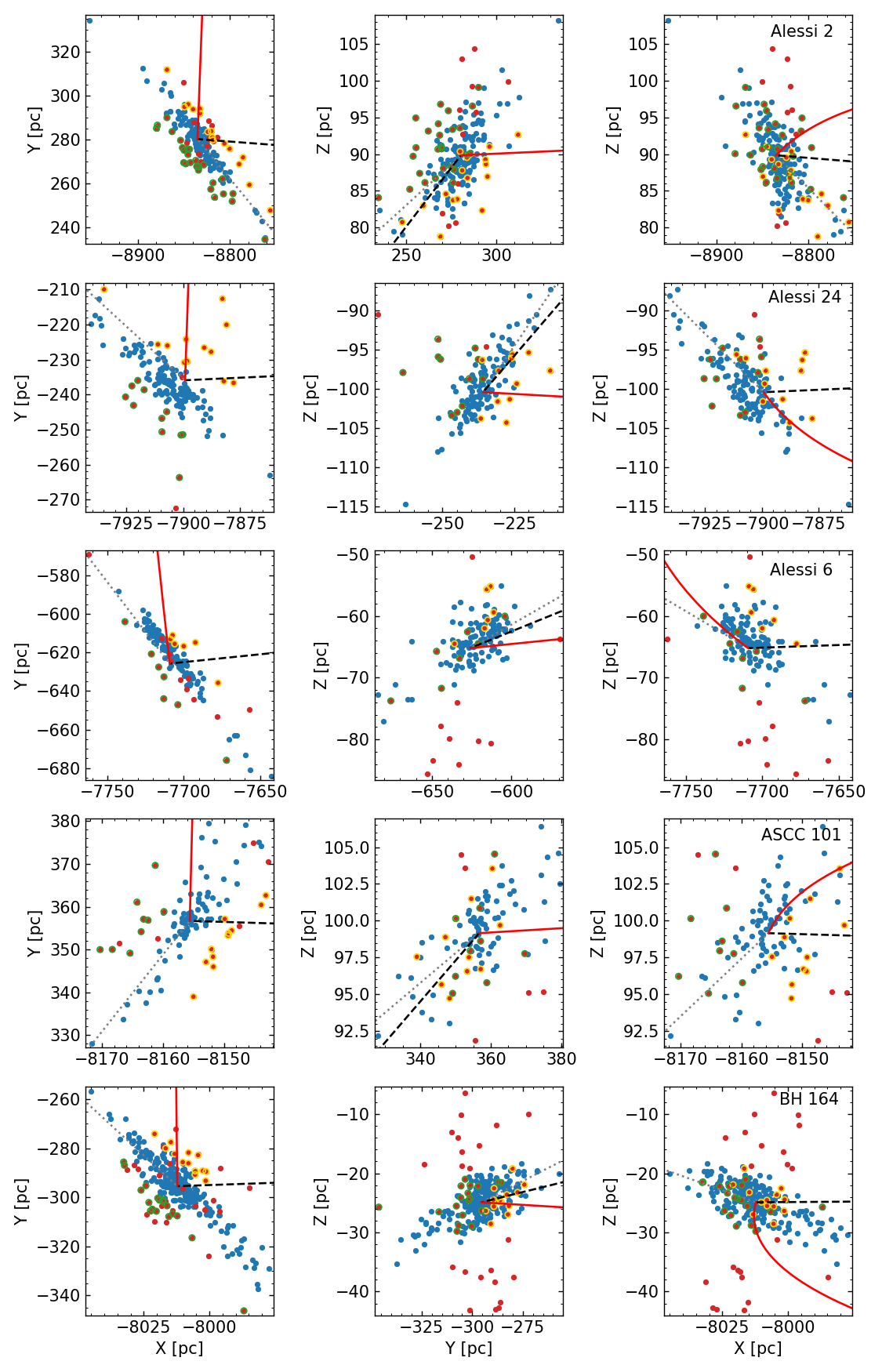}
    \caption{Same as Figure~\ref{fig:orbit} but for the other open clusters with tidal tails.}
    \label{fig:orbit_append}
\end{figure*}

\begin{figure*}\ContinuedFloat
	\includegraphics[width=0.83\textwidth]{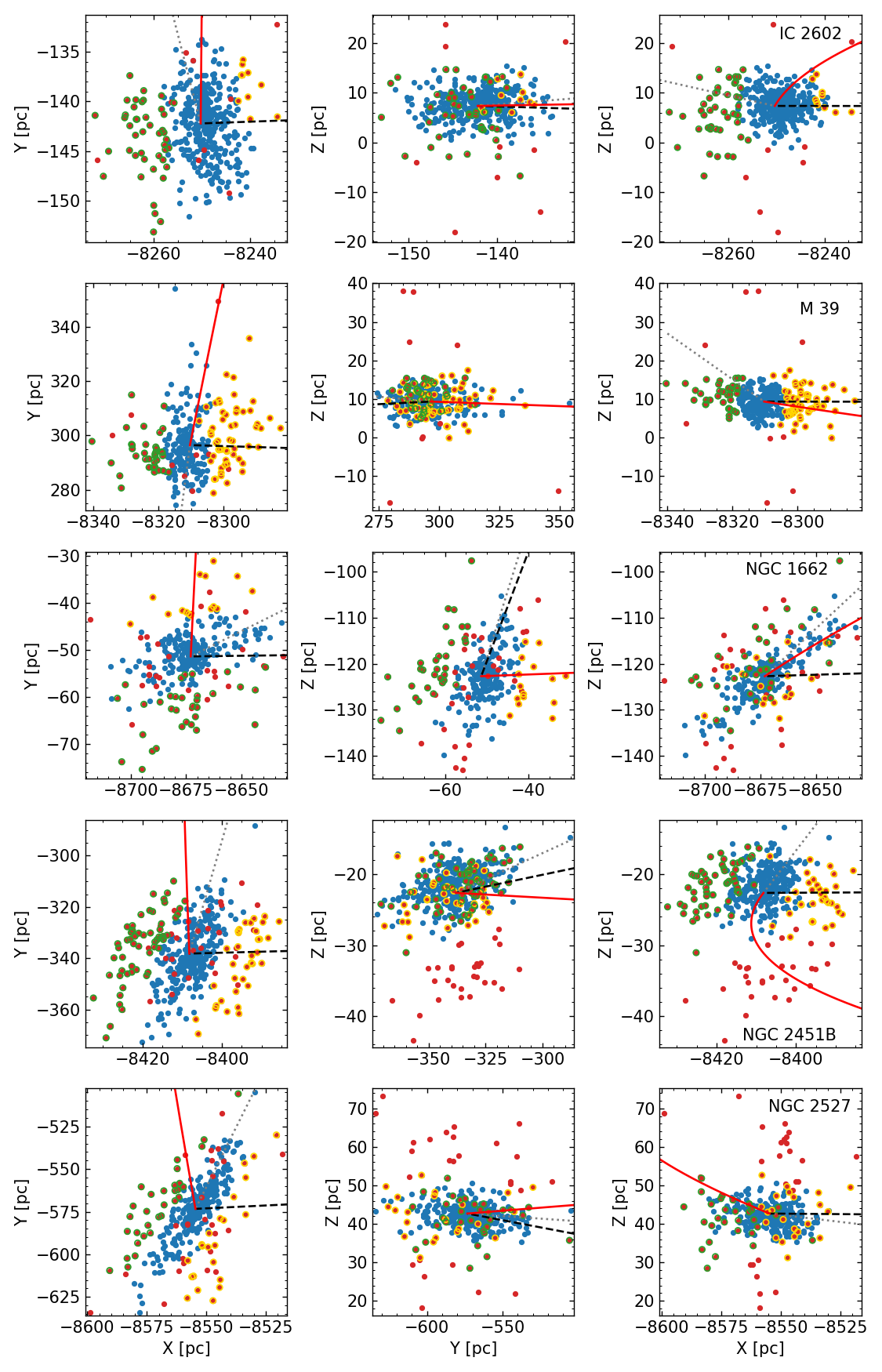}
    \caption{Same as Figure~\ref{fig:orbit} but for the other open clusters with tidal tails.}
\end{figure*}

\begin{figure*}\ContinuedFloat
	\includegraphics[width=0.83\textwidth]{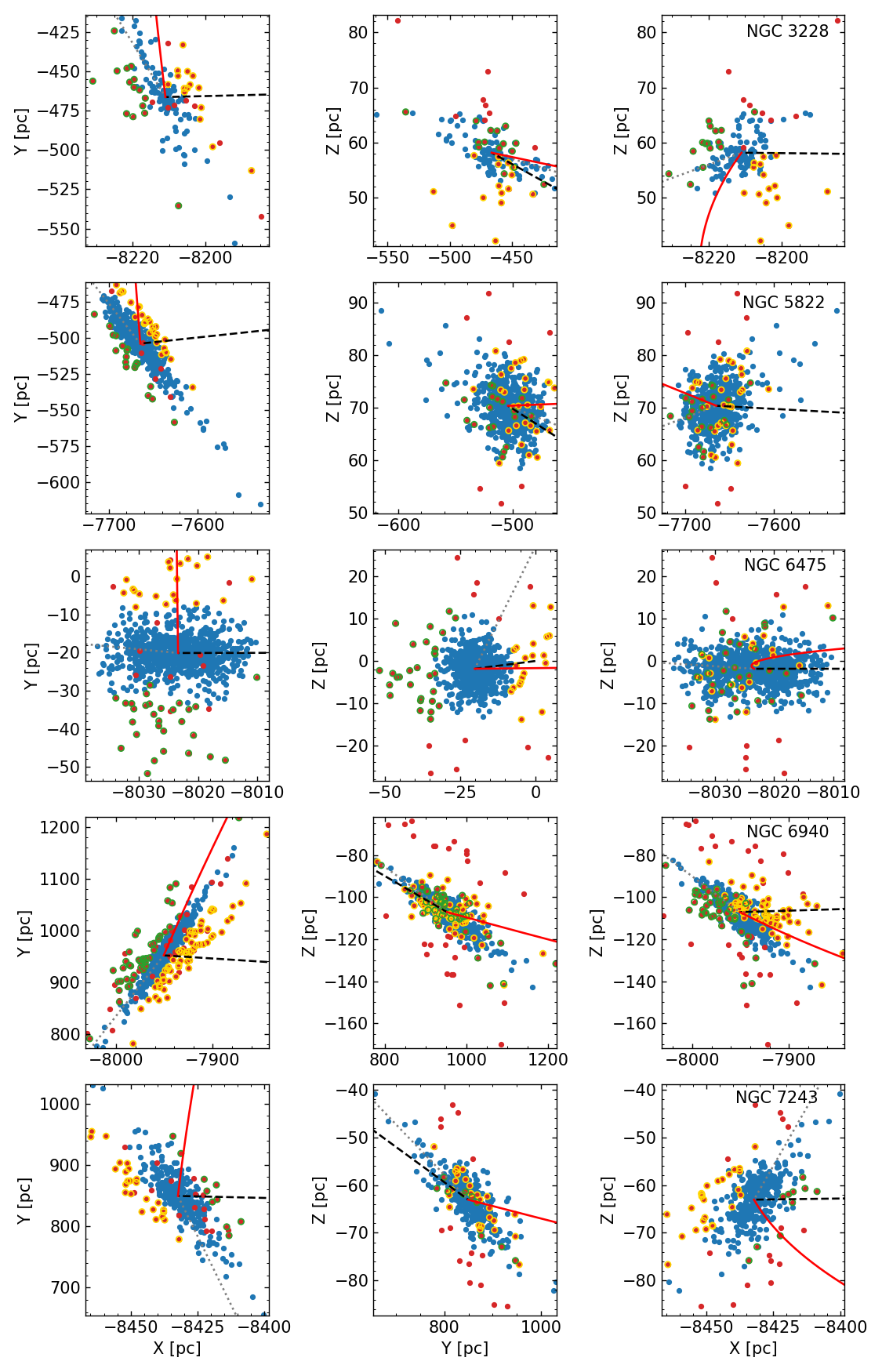}
    \caption{Same as Figure~\ref{fig:orbit} but for the other open clusters with tidal tails.}
\end{figure*}

\begin{figure*}\ContinuedFloat
	\includegraphics[width=0.83\textwidth]{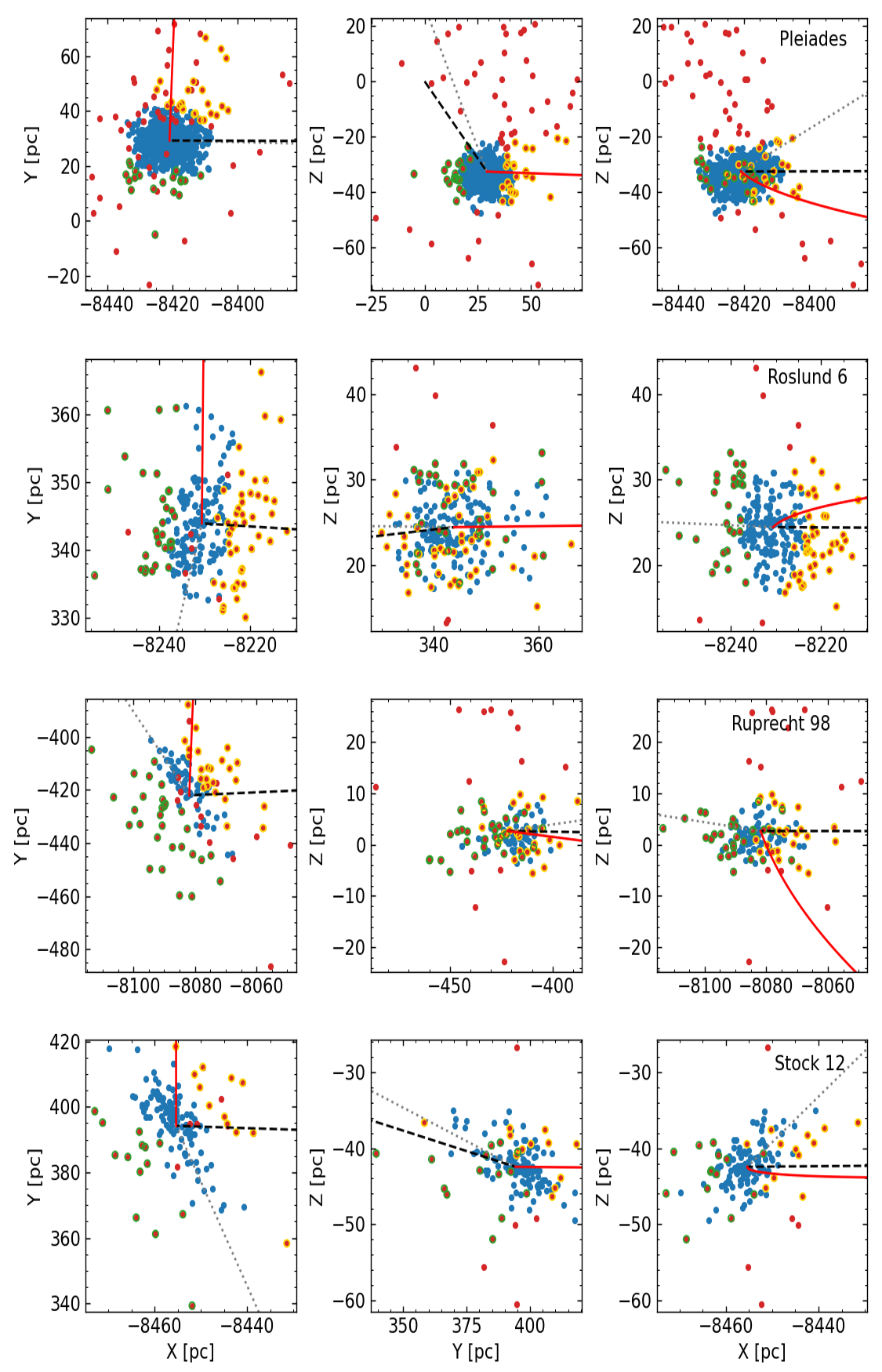}
    \caption{Same as Figure~\ref{fig:orbit} but for the other open clusters with tidal tails.}
\end{figure*}

\bsp	
\label{lastpage}
\end{document}